\documentclass[letterpaper,11pt,onecolumn,draftcls]{IEEEtran}

\usepackage{amsmath,amssymb,amsthm}
\usepackage{graphicx}
\usepackage{bm}
\usepackage{tikz}

\newtheorem{theorem}{Theorem}
\newtheorem{corollary}{Corollary}
\newtheorem{lemma}{Lemma}

\theoremstyle{definition}
\newtheorem{definition}{Definition}

\theoremstyle{remark}
\newtheorem{remark}{Remark}

\def\e{\mathop{\mathrm{e}}\nolimits}
\newcommand{\bbset}[1]{\mathbb{#1}}
\newcommand{\N}[0]{\bbset{N}}      
\newcommand{\Z}[0]{\bbset{Z}}      
\newcommand{\R}[0]{\bbset{R}}      
\newcommand{\C}[0]{\bbset{C}}      
\newcommand{\Bs}[0]{\mathcal{B}}      
\newcommand{\PWs}[0]{\mathcal{PW}}    
\newcommand{\Hto}[0]{H}               
\newcommand{\Ht}[1]{\tilde{#1}}       
\newcommand{\Ft}[1]{\hat{#1}}         
\newcommand{\iu}[0]{i}                
\newcommand{\convolution}[0]{*}       
\newcommand{\VPint}[0]{\, \text{V.P.}\! \int}   
\newcommand{\defequal}[0]{\mathrel{\mathop{:}}=}  
\newcommand{\di}[1]{\;\mathrm{d}#1}
\newcommand{\spacedot}[0]{\,\cdot\,}      
\DeclareMathOperator{\sgn}{sgn}           
\DeclareMathOperator{\sinc}{sinc}         
\newcommand{\timeinterval}[0]{\tau}

\begin{document}

\title{Strong Divergence for System Approximations%
  \thanks{H. Boche was partly supported by the German Research Foundation (DFG) under grant BO
    1734/22-1. U. M{\"o}nich was supported by the German Research Foundation (DFG)
    under grant MO 2572/1-1.}}

\author{Holger~Boche~and~Ullrich~J.~M{\"o}nich
  \thanks{Holger Boche is with the Technische Universit\"at M\"unchen,
    Lehrstuhl f\"ur Theoretische Informationstechnik, Germany (e-mail: boche@tum.de);
    Ullrich J. M\"onich is with the Massachusetts Institute of Technology,
    Research Laboratory of Electronics, USA (e-mail: moenich@mit.edu).}%
  \thanks{Preprint accepted for publication in Problems of Information Transmission. The material in this paper was presented in part at the 2015
    IEEE International Conference on Acoustics, Speech, and Signal Processing.}
}

\markboth{Problems of Information Transmission}%
{Boche and M\"onich}

\maketitle

\begin{abstract}
  In this paper we analyze the approximation of stable linear time-invariant systems, like
  the Hilbert transform, by sampling series for bandlimited functions in the Paley--Wiener
  space $\PWs_{\pi}^{1}$. 
  It is known that there exist systems and functions such that the approximation process
  is weakly divergent, i.e., divergent for certain subsequences. 
  Here we strengthen this result by proving strong divergence, i.e., divergence for all
  subsequences. 
  Further, in case of divergence, we give the divergence speed. 
  We consider sampling at Nyquist rate as well as oversampling with adaptive choice of the
  kernel. 
  Finally, connections between strong divergence and the Banach--Steinhaus theorem, which
  is not powerful enough to prove strong divergence, are discussed.
\end{abstract}


\section{Introduction}\label{sec:intro}

Sampling theory studies the reconstruction of a function in terms of its samples. 
In addition to its mathematical significance, sampling theory plays a fundamental role in
modern signal and information processing because it is the basis for today's digital world
\cite{shannon49}.

The fundamental initial result of the theory states that the Shannon sampling series
\begin{equation}\label{eq:shannon_series}
  \sum_{k=-\infty}^{\infty} f(k) \frac{\sin(\pi(t-k))}{\pi(t-k)}
\end{equation}
can be used to reconstruct bandlimited functions $f$ with finite $L^2$-norm from their
samples $\{f(k)\}_{k \in \Z}$. 
Since this initial result, many different sampling theorems have been developed, and
determining the function classes for which the theorems hold and the mode of convergence
now constitute an entire area of research
\cite{jerri77,higgins85,butzer88,marvasti01_book}.

In this paper we study the convergence behavior of different sampling series for the
Paley--Wiener space $\PWs_{\pi}^{1}$ consisting of absolutely integrable bandlimited
functions. 
Analyzing sampling series and finding sampling theorems for the Paley--Wiener space
$\PWs_{\pi}^{1}$ has a long tradition \cite{brown67,butzer88,butzer92}. 
Since Shannon's initial result for $\PWs_{\pi}^{2}$ \cite{shannon49}, efforts have been
made to extend it to larger signal spaces \cite{brown67,higgins96_book,butzer11}.

In this paper we prove strong divergence, i.e., divergence for all subsequences, for
different sampling series, where only weak divergence, i.e., divergence for certain
subsequences, was known before, and further, we give the order of divergence. 
We also study the approximation of linear time-invariant (LTI) systems and show that we
have strong divergence there, even in the case of oversampling. 
Interestingly, it is possible to show strong divergence if the system is the Hilbert
transform, which is a stable LTI system for $\PWs_{\pi}^{1}$, i.e. 
the space under consideration.

In addition to the specific questions about the convergence and divergence behavior of
sampling series, there also rises a general mathematical question in the context of the
analyses in this paper: can we develop universal mathematical techniques for the
convergence and divergence analysis of adaptive signal processing procedures? 
For example, the Banach--Steinhaus theory from functional analysis can be seen as a
mathematical tool for analyzing non-adaptive signal processing procedures. 
The question is whether a similar theory can also be developed for adaptive signal
processing.

In the next section we will introduce some notation and then, in
Section~\ref{sec:problem-formulation}, we will give a more detailed
motivation of the problem.

\section{Notation}\label{sec:notation}

Let $\Ft{f}$ denote the Fourier transform of a function $f$, where $\Ft{f}$ is to be
understood in the distributional sense. 
By $L^p(\R)$, $1\leq p \leq \infty$, we denote the usual $L^p$-spaces, equipped with the
norm $\lVert\spacedot\rVert_p$.

For $\sigma > 0$ let $\Bs_{\sigma}$ be the set of all entire functions $f$ with the
property that for all $\epsilon > 0$ there exists a constant $C(\epsilon)$ with
$\lvert f(z) \rvert \leq C(\epsilon) \exp \bigl( (\sigma+\epsilon) \lvert z \rvert \bigr)$
for all $z \in \C$. 
The Bernstein space $\Bs_{\sigma}^{p}$ consists of all functions in $\Bs_{\sigma}$ whose
restriction to the real line is in $L^p(\R)$, $1 \leq p \leq \infty$. 
A function in $\Bs_{\sigma}^{p}$ is called bandlimited to $\sigma$.

For $\sigma>0$ and $1\leq p\leq \infty$, we denote by $\PWs_{\sigma}^{p}$ the Paley-Wiener
space of functions $f$ with a representation
$f(z)=1/(2\pi) \int_{-\sigma}^{\sigma} g(\omega) \e^{\iu z\omega} \di{\omega}$,
$z \in \C$, for some $g \in L^p[-\sigma,\sigma]$. 
If $f\in \PWs_{\sigma}^{p}$, then $g(\omega)=\Ft{f}(\omega)$. 
The norm for $\PWs_{\sigma}^{p}$, $1\leq p < \infty$, is given by
$\lVert f \rVert_{\PWs_{\sigma}^{p}}=( 1/(2\pi)\int_{-\sigma}^{\sigma} \lvert
\Ft{f}(\omega) \rvert^p \di{\omega} )^{1/p}$.

\section{Problem Formulation and Discussion}\label{sec:problem-formulation}

\subsection{Adaptive Function Reconstruction}\label{sec:adapt-funct-reconstr}
Before we state our main results, we present, motivate, and discuss the problems and main
questions that we treat in this paper. 
Let
\begin{equation}\label{def_SN}
  (S_N f)(t)
  \defequal
  \sum_{k=-N}^{N} f(k) \frac{\sin(\pi(t-k))}{\pi(t-k)} 
\end{equation}
denote the finite Shannon sampling series. 
It is well-known that $S_N f$ converges locally uniformly to $f$ for all functions
$f \in \PWs_{\pi}^{1}$ as $N$ tends to infinity \cite{brown67,butzer88,butzer92}. 
However, the series is not globally uniformly convergent. 
The quantity
\begin{equation*}
  P_N f
  \defequal
  \max_{t \in \R}  \left| f(t) - (S_N f)(t) \right| ,
\end{equation*}
i.e., the peak value of the reconstruction error, diverges for certain
$f \in \PWs_{\pi}^{1}$ as $N$ tends to infinity. 
In \cite{boche11a} it has been shown that there exists a function $f \in \PWs_{\pi}^{1}$
such that
\begin{equation}\label{eq:limsup_divergence}
\limsup_{N \rightarrow \infty} P_N f = \infty .
\end{equation}
Since the uniform boundedness theorem has been applied in the proof of
\eqref{eq:limsup_divergence}, it follows immediately that the set of functions
$\mathcal{D} \subset \PWs_{\pi}^{1}$, for which \eqref{eq:limsup_divergence} holds, is a
residual set.

However, the divergence is only given in terms of the $\limsup$. 
In a sense this is a weak notion of divergence, because it merely guarantees the existence
of a subsequence $\{N_n\}_{n \in \N}$ of the natural numbers such that
$\lim_{n \rightarrow \infty} P_{N_n} f = \infty$ for a certain $f \in D$. 
This leaves the possibility that there is a different subsequence $\{N_n^*\}_{n \in \N}$
such that $\lim_{n \rightarrow \infty} P_{N_n^*} f = 0$.

This possibility was discussed in \cite{boche14e_accepted}, and two conceivable situations
were phrased in two questions.

\noindent \textbf{Question Q1:}~\\ Does there, for every $f \in \PWs_{\pi}^{1}$, exist a subsequence $\{N_n\}_{n  \in \N} = \{N_n(f)\}_{n \in \N}$ of the natural numbers such that $\sup_{n \in \N} P_{N_n} f < \infty$?

\noindent \textbf{Question Q2:}~\\ Does there exist a subsequence $\{N_n\}_{n
  \in \N}$ of the natural numbers such that $\sup_{n \in \N} P_{N_n} f < \infty$
for all $f \in \PWs_{\pi}^{1}$?
\vspace{1em}

\noindent Note that the subsequence $\{N_n(f)\}_{n \in \N}$ in Question Q1 can depend on
the function $f$ that shall be reconstructed. 
Thus, the reconstruction process $S_{N_n(f)}$ is adapted to the function $f$. 
The problem of finding an index sequence, depending on the function $f$, that is suitable
for achieving the desired goal, is the task of adaptive signal processing. 
In our case it is the adaptive reconstruction of $f$ from measurement values. 
Adaptive signal processing covers most of the practical important applications.

In contrast, the subsequence $\{N_n\}_{n \in \N}$ in Question Q2 is universal in the sense
that it does not depend on $f$. 
Obviously, a positive answer to Question Q2 implies a positive answer to Question Q1.

This brings us to the notion of strong divergence. 
We say that a sequence $\{a_n\}_{n \in \N} \subset \R$ diverges strongly if
$\lim_{n \rightarrow \infty} \lvert a_n \rvert = \infty$. 
Clearly this is a stronger statement than
$\limsup_{n \rightarrow \infty} \lvert a_n \rvert = \infty$, because in case of strong
divergence we have $\lim_{n \rightarrow \infty} \lvert a_{N_n} \rvert = \infty$ for all
subsequences $\{N_n\}_{n \in \N}$ of the natural numbers.

So, if $P_N f$ is strongly divergent for all $f \in \PWs_{\pi}^{1}$, then Question Q1 and
consequently Question Q2 have to be answered in the negative.

Divergence results as in \eqref{eq:limsup_divergence} are usually proved by using the
uniform boundedness principle, which is also known as Banach--Steinhaus theorem
\cite{banach27}. 
As an immediate consequence, the obtained divergence is in terms of the $\limsup$ and not
a statement about strong divergence. 
However, the strength of the uniform boundedness principle is that the divergence
statement holds not only for a single function but immediately for a large set of
functions: the set of all functions for which we have divergence is a residual set.

Since the publication of Banach and Steinhaus \cite{banach27,banach31}, the
Banach--Stein\-haus theory has been developed further and has today become an important
part of functional analysis. 
There also have been efforts to extend the Banach--Steinhaus theory into different
directions \cite{stein61,dickmeis85,dickmeis85a,velasco95,jachymski05}. 
However, these extensions do not cover Question Q1, that is, they provide no tools to
analyze adaptive signal processing techniques in the sense of Question Q1. 
Next, we will further discuss Question Q1 and the difference to the Banach--Steinhaus
theory.

It is tempting to try to use the uniform boundedness principle to prove that the answer to
Question Q1 is no. 
Let $N=\{N_n\}_{n \in \N}$ be a subsequence of the natural numbers. 
Then the uniform boundedness principle states the existence of a residual set
$\mathcal{D}(N) \subset \PWs_{\pi}^{1}$ such that
\begin{equation*}
  \limsup_{n \rightarrow \infty} P_{N_n} f = \infty
\end{equation*}
for all $f \in \mathcal{D}(N)$. If we could prove that
\begin{equation*}
  \bigcap_{\text{$N$ is a subsequence of $\N$}} \mathcal{D}(N)
  \neq
  \varnothing ,
\end{equation*}
then the answer to Question Q1 would be no. 
However, the set of all subsequences of $\N$ contains uncountably many elements, and the
uncountable intersection of residuals set may be empty. 
Hence, we cannot use this approach to prove strong divergence. 
In Section~\ref{sec:discussion1} we will see an example where we have this situation.

In \cite{boche14c} it has been proved, using a different proof technique, that there
exists a function $f \in \PWs_{\pi}^{1}$ such that $P_N f$ diverges strongly, i.e., that
$\lim_{N \rightarrow \infty} P_N f = \infty$. 
Hence, neither Question Q1 nor Question Q2 can be answered in the affirmative for the
Shannon sampling series. 
Moreover in \cite{boche14c}, the authors posed a question about the divergence speed of
$P_N f$ that we will answer in Section~\ref{sec:behav-conj-shann}.

It is interesting to note that the application of the uniform boundedness principle does
not require a deep analysis of the approximation process $S_N f$. 
A simple evaluation of the operator norm
\begin{equation*}
  \lVert S_N \rVert = \sup_{\substack{f \in \PWs_{\pi}^{1}, \\ \lVert f
    \rVert_{\PWs_{\pi}^{1}} = 1}} \lVert S_N f \rVert_{\infty}
\end{equation*}
is sufficient.

It would be desirable to have a theorem, analogous to the uniform boundedness theorem,
that can be used to prove strong divergences. 
Currently, little is known about the structure of this problem, and it is unclear
whether such a theorem can exist \cite{banach31,velasco95,jachymski05}. 
Due to the lack of such a theory, we need to develop proof strategies which are tailored
to the specific situation of the different approximation processes in order to show
strong divergence.

After publication of \cite{boche14c}, the first author noticed that Paul Erd{\H{o}}s
analyzed similar questions for the Lagrange interpolation on Chebyshev nodes
\cite{erdos41}. 
However, in \cite{erdos43} Erd{\H{o}}s observed that his own proof was erroneous, and he
was not able to present a correct proof. 
It seems that the original problem is still open.

\subsection{System Approximation}\label{sec:system-approximation}

A more general problem than the reconstruction problem, where the goal is to reconstruct a
bandlimited functions $f$ from its samples $\{f(k)\}_{k \in \Z}$, is the system
approximation problem, where the goal is to approximate the output $Tf$ of a stable LTI
system $T$ from the samples $\{f(k)\}_{k \in \Z}$ of the input function $f$. 
This is the situation that is encountered in digital signal processing applications, where
the interest is not in the reconstruction of a signal, but rather in the implementation of
a system, i.e, the interest is in some transformation $Tf$ of the sampled input signal
$f$. 
For discussions of the significance of signal processing as the basis of our digital
information age, see for example \cite{boche14d} and references therein.

We briefly review some basic definitions and facts about stable linear time-invariant
(LTI) systems.

A linear system $T:\PWs_{\pi}^{p} \rightarrow \PWs_{\pi}^{p}$, $1 \leq p \leq \infty$, is
called stable if the operator $T$ is bounded, i.e., if
$\lVert T \rVert = \sup_{\lVert f \rVert_{\PWs_{\pi}^{p}}\leq 1} \lVert Tf
\rVert_{\PWs_{\pi}^{p}} < \infty$. 
Furthermore, it is called time-invariant if $(Tf(\spacedot - a))(t)=(Tf)(t-a)$ for all
$f \in \PWs_{\pi}^{p}$ and $t,a \in \R$. 
For every stable LTI system $T:\PWs_{\pi}^{1} \rightarrow \PWs_{\pi}^{1}$, there exists
exactly one function $\Ft{h}_T \in L^\infty[-\pi,\pi]$ such that
\begin{equation}\label{eq:frequency_domain_representation}
  (Tf)(t)=\frac{1}{2\pi} \int_{-\pi}^{\pi} 
  \Ft{f}(\omega) \Ft{h}_T(\omega) \e^{\iu \omega t} \di{\omega} ,\quad t \in \R,
\end{equation}
for all $f \in \PWs_{\pi}^{1}$ \cite{boche10g}. 
Conversely, every function $\Ft{h}_T \in L^\infty[-\pi,\pi]$ defines a stable LTI system
$T:\PWs_{\pi}^{1} \rightarrow \PWs_{\pi}^{1}$. 
The operator norm of a stable LTI system $T$ is given by
$\lVert T \rVert=\lVert \Ft{h} \rVert_{L^\infty[-\pi,\pi]}$. 
Furthermore, it can be shown that the representation
\eqref{eq:frequency_domain_representation} with $\Ft{h}_T \in L^\infty[-\pi,\pi]$ is also
valid for all stable LTI systems $T: \PWs_{\pi}^{2} \rightarrow \PWs_{\pi}^{2}$. 
Therefore, every stable LTI system that maps $\PWs_{\pi}^{1}$ in $\PWs_{\pi}^{1}$ maps
$\PWs_{\pi}^{2}$ in $\PWs_{\pi}^{2}$, and vice versa. 
Note that $\Ft{h}_T \in L^\infty[-\pi,\pi] \subset L^2[-\pi,\pi]$, and consequently
$h_T \in \PWs_{\pi}^{2}$.

Similar to the Shannon sampling series \eqref{eq:shannon_series}, which was used in the function reconstruction problem, we can use the approximation process
\begin{equation}\label{eq:system_approximation_process}
  \sum_{k=-\infty}^{\infty} f(k) h_T(t-k)
\end{equation}
in the system approximation problem. In order to analyze the convergence behavior of \eqref{eq:system_approximation_process}, we introduce the abbreviation
\begin{equation}\label{eq:def_TN}
  (T_Nf)(t)
  \defequal
  \sum_{k=-N}^{N} f(k) h_T(t-k) .
\end{equation}

As already mentioned before, for certain functions in $f \in \PWs_{\pi}^{1}$, the peak
value of the reconstruction process $\lVert S_N f \rVert_{\infty}$ diverges strongly as
$N$ tends to infinity. 
However, in the case of oversampling, i.e., the case where the sampling rate is higher
than Nyquist rate, the function reconstruction process $S_N f$ converges globally
uniformly \cite{boche09f}. 
This is a situation where oversampling helps improve the convergence behavior, consistent
with engineering intuition. 
In contrast, the convergence behavior of the system approximation process
\eqref{eq:system_approximation_process} does not improve with oversampling
\cite{boche10g}: for every $t \in \R$ and every $\sigma \in (0,\pi]$ there exist stable
LTI systems $T \colon \PWs_{\pi}^{1} \to \PWs_{\pi}^{1}$ and functions
$f \in \PWs_{\sigma}^{1}$ such that
\begin{equation*}
  \limsup_{N \rightarrow \infty} \lvert (Tf)(t) - (T_N f)(t) \rvert 
  =
  \infty .
\end{equation*}

In this paper we want to refine the Questions Q1 and Q2 and analyze five questions:
\begin{enumerate}
\item Do we have the same strong divergence for the system approximation process $T_N f$?
\item Is it possible to obtain quantitative results about the divergence speed?
\item What happens in the case of oversampling?
\item What are the cases where no strong divergence can occur, and how can they be characterized?
\item How large is the set of functions with strong divergence?
\end{enumerate}

We will treat the fifth question only briefly in Section~\ref{sec:discussion}, where we
present one example where the set of functions with strong divergence is empty and two
examples where this set is a residual set. 
In general, the answer to this question is unknown.

\section{Behavior of the Conjugated Shannon Sampling Series and the Shannon Sampling Series}\label{sec:behav-conj-shann}

In this section we analyze the behavior of conjugated Shannon sampling series and the
Shannon sampling series. 
We first study the conjugated Shannon sampling series with critical sampling at Nyquist
rate, i.e., the case without oversampling, and show that the answer to Question Q1 is
negative in this case. 
To this end, let $S_N f$ denote the finite Shannon sampling series as defined in
\eqref{def_SN}, and
\begin{equation}\label{eq:finite_conj_shannon}
  (\Hto_N f)(t)
  \defequal
  (\Hto S_N f)(t)
  =
  \sum_{k=-N}^{N} f(k) \frac{1-\cos(\pi(t-k))}{\pi(t-k)}
\end{equation}
the conjugated finite Shannon sampling series. 
$\Hto$ denotes the Hilbert transform which is defined as the principal value integral
\begin{align}
  (\Hto f)(t)
  &=
  \frac{1}{\pi} \VPint_{-\infty}^{\infty} \frac{f(\tau)}{t-\tau} \di{\tau} 
  =
  \frac{1}{\pi} \lim_{\epsilon \rightarrow 0} \int\limits_{\epsilon \leq \lvert t-\tau \rvert \leq
    \frac{1}{\epsilon}} \frac{f(\tau)}{t-\tau} \di{\tau} \notag .
  \label{eq:def_hilbert_transformation_introduction}
\end{align}
The Hilbert transform is of enormous practical significance and plays a central role in
the analysis of signal properties \cite{fink66,korzhik69,vakman72,gabor46,huang98,huang99,huang03}. 
For further applications, see for example \cite{pohl09_book} and references therein.

It is well-known that $H_N f$ converges locally uniformly to $Hf$ as $N$ tends to
infinity, that is, for $\timeinterval > 0$ we have
\begin{equation*}
  \lim_{N \rightarrow \infty} \left(
    \max_{\lvert t \rvert \leq \timeinterval}
    \lvert (\Hto f)(t) - (\Hto_N f)(t) \rvert
  \right)
  =
  0 .
\end{equation*}
The next theorem gives an answer about the global behavior of
\eqref{eq:finite_conj_shannon}.
\begin{theorem}\label{th:divergence_speed_conj_shannon_no_oversampl}
  Let $\{\epsilon_N\}_{N \in \N}$ be an arbitrary sequence of positive numbers converging to zero.
  There exists a function $f_1 \in \PWs_{\pi}^{1}$ such that
  \begin{equation*}
    \lim_{N \rightarrow \infty} \frac{1}{\epsilon_N \log(N)} \left(
      \max_{t \in \R} \left(
        \sum_{k=-N}^{N} f_1(k) \frac{1-\cos(\pi(t-k))}{\pi(t-k)}
      \right)
    \right)
    =
    \infty
  \end{equation*}
  and
  \begin{equation*}
    \lim_{N \rightarrow \infty} \frac{1}{\epsilon_N \log(N)} \left(
      \min_{t \in \R} \left(
        \sum_{k=-N}^{N} f_1(k) \frac{1-\cos(\pi(t-k))}{\pi(t-k)}
      \right)
    \right)
    =
    -\infty .
  \end{equation*}
\end{theorem}

\begin{IEEEproof}
  Let $\{\epsilon_N\}_{N \in \N}$ be an arbitrary sequence of positive numbers converging
  to zero, and $\bar{\epsilon}_N = \max_{M \geq N} \epsilon_M$, $N \in \N$. 
  Note that $\bar{\epsilon}_N \geq \epsilon_N$ for all $N \in \N$. 
  Further, let $\{N_k\}_{k \in \N}$ be a strictly monotonically increasing sequence of
  natural numbers, such that $\bar{\epsilon}_{N_k} > \bar{\epsilon}_{N_{k+1}}$,
  $k \in \N$. 
  We set $\delta_k = \sqrt{\bar{\epsilon}_{N_k}} - \sqrt{\bar{\epsilon}_{N_{k+1}}}$,
  $k \in \N$. 
  It follows that $\delta_k > 0$ for all $k \in \N$ and that
  \begin{equation}\label{eq:pr:th:divergence_speed_conj_shannon_no_oversampl:sum_delta_k}
    \sum_{k=1}^{\infty} \delta_k = \sqrt{\bar{\epsilon}_{N_1}} < \infty .
  \end{equation}
  For $N \in \N$ we define the functions
  \begin{equation*}
    w_N(t)
    =
    \sum_{k=-\infty}^{\infty} w_N(k)
    \frac{\sin(\pi(t-k))}{\pi(t-k)}, \quad t \in \R,
  \end{equation*}
  where $w_N(k)$ is given by
  \begin{equation*}
    w_N(k)=\begin{cases}
      1, & \lvert k \rvert \leq N, \\
      1-\frac{\lvert k \rvert - N}{N},  & N < \lvert k \rvert < 2N,\\
      0,  & \lvert k \rvert \geq 2N .
    \end{cases}
  \end{equation*}
  Note that we have $w_N \in \PWs_{\pi}^{1}$ and $\lVert w_N \rVert_{\PWs_{\pi}^{1}} < 3$
  for all $N \in \N$ \cite{boche10d}. 
  Based on $w_N$ we define function
  \begin{equation}\label{eq:pr:th:divergence_speed_conj_shannon_no_oversampl:def_f1_wk}
    f_1
    =
    \sum_{k=1}^{\infty} \delta_k w_{N_{k+1}} .
  \end{equation}
  Since $\lVert \delta_{k} w_{N_{k+1}} \rVert_{\PWs_{\pi}^{1}} < 3 \delta_{k}$ and because
  of \eqref{eq:pr:th:divergence_speed_conj_shannon_no_oversampl:sum_delta_k}, it follows
  that the partial sums of the series in
  \eqref{eq:pr:th:divergence_speed_conj_shannon_no_oversampl:def_f1_wk} form a Cauchy
  sequence in $\PWs_{\pi}^{1}$, and thus the series in
  \eqref{eq:pr:th:divergence_speed_conj_shannon_no_oversampl:def_f1_wk} converges in the
  $\PWs_{\pi}^{1}$-norm and consequently uniformly on $\R$. 
  Let $N \in \N$ be arbitrary but fixed. 
  For $t_N=N+1$, it follows that
  \begin{align}
    \sum_{l=-N}^{N} f_1(l) \frac{1-\cos(\pi (t_N-l))}{\pi (t_N-l)}
    &=
    \sum_{l=-N}^{N} f_1(l) \frac{1- (-1)^{N+1-l}}{\pi (N+1-l)} . \label{eq:pr:th:divergence_speed_conj_shannon_no_oversampl:1}
  \end{align}
  There exists exactly one $\hat{k} \in \N$ such that
  $N \in [N_{\hat{k}}, N_{\hat{k}+1})$. 
  We have
  \begin{align}
    \sum_{l=-N}^{N} f_1(l) \frac{1- (-1)^{N+1-l}}{\pi (N+1-l)}
    &=
    \sum_{k=1}^{\infty} \frac{\delta_k}{\pi} \sum_{l=-N}^{N}
    w_{N_{k+1}}(l)\frac{1- (-1)^{N+1-l}}{N+1-l} \notag \\
    &\geq
    \sum_{k=\hat{k}}^{\infty} \frac{\delta_k}{\pi} \sum_{l=-N}^{N}
    w_{N_{k+1}}(l)\frac{1- (-1)^{N+1-l}}{N+1-l} \notag \\
    &=
    \sum_{k=\hat{k}}^{\infty} \frac{\delta_k}{\pi} \sum_{l=-N}^{N}
    \frac{1- (-1)^{N+1-l}}{N+1-l} , \label{eq:pr:th:divergence_speed_conj_shannon_no_oversampl:2}
  \end{align}
  where we used that $w_{N_{k+1}}(l) = 1$ for all $k \geq \hat{k}$ and all
  $\lvert l \rvert \leq N$.
  Further, we have
  \begin{align}
    \sum_{k=\hat{k}}^{\infty} \frac{\delta_k}{\pi} \sum_{l=-N}^{N}
    \frac{1- (-1)^{N+1-l}}{N+1-l}
    &=
    \frac{1}{\pi} \sum_{k=\hat{k}}^{\infty} \delta_k \sum_{l=1}^{2N+1}
    \frac{1- (-1)^{l}}{l} \notag \\
    &=
    \frac{1}{\pi} \sum_{k=\hat{k}}^{\infty} \delta_k \sum_{l=0}^{N}
    \frac{2}{2l+1} \notag \\
    &\geq
    \frac{1}{\pi} \log(2N+3) \sum_{k=\hat{k}}^{\infty} \delta_k \notag \\
    &=
    \frac{1}{\pi} \log(2N+3) \sqrt{\bar{\epsilon}_{N_{\hat{k}}}} \notag \\
    &\geq
    \frac{1}{\pi} \epsilon_N \log(N) \frac{1}{\sqrt{\epsilon_{N}}} \label{eq:pr:th:divergence_speed_conj_shannon_no_oversampl:3}
  \end{align}
  because $N \geq N_{\hat{k}}$ and thus
  $\sqrt{\bar{\epsilon}_{N_{\hat{k}}}} \geq \sqrt{\epsilon_{N_{\hat{k}}}} \geq
  \sqrt{\epsilon_{N}}$. 
  From \eqref{eq:pr:th:divergence_speed_conj_shannon_no_oversampl:1}--\eqref{eq:pr:th:divergence_speed_conj_shannon_no_oversampl:3},
  we see that
  \begin{equation*}
    \sum_{l=-N}^{N} f_1(l) \frac{1-\cos(\pi (t_N-l))}{\pi (t_N-l)}
    \geq
    \frac{1}{\pi} \epsilon_N \log(N) \frac{1}{\sqrt{\epsilon_{N}}}
  \end{equation*}
  for all $N \in \N$, which in turn implies that
  \begin{equation*}
    \lim_{N \rightarrow \infty}
    \frac{1}{\epsilon_N \log(N)}
    \left( \max_{t \in \R} \left(
        \sum_{k=-N}^{N} f_1(k) \frac{1-\cos(\pi(t-k))}{\pi(t-k)}
      \right) \right)
    =
    \infty .
  \end{equation*}
  The second assertion
  \begin{equation*}
    \lim_{N \rightarrow \infty}
    \frac{1}{\epsilon_N \log(N)}
    \left( \min_{t \in \R} \left(
        \sum_{k=-N}^{N} f_1(k) \frac{1-\cos(\pi(t-k))}{\pi(t-k)}
      \right) \right)
    =
    -\infty 
  \end{equation*}
  is proved by choosing $t_N = -N-1$ instead of $t_N = N+1$.
\end{IEEEproof}

Next, we analyze the oversampling case for the conjugated Shannon sampling series, i.e.,
we treat question 3 from Section~\ref{sec:system-approximation}.

For the Shannon sampling series the convergence behavior in the case of oversampling is
clear: we have global uniform convergence \cite{boche09f}. 
However, this is not true for the conjugated Shannon sampling series as the next theorem
shows.
\begin{theorem}\label{th:divergence_speed_conj_shannon_oversampl}
  Let $\{\epsilon_N\}_{N \in \N}$ be an arbitrary sequence of positive numbers converging
  to zero.
  For every $\sigma \in (0,\pi]$ there exists a function $f_\sigma \in \PWs_{\sigma}^{1}$
  such that
  \begin{equation*}
    \lim_{N \rightarrow \infty} \frac{1}{\epsilon_N \log(N)} \left(
      \max_{t \in \R} \left(
        \sum_{k=-N}^{N} f_\sigma(k) \frac{1-\cos(\pi(t-k))}{\pi(t-k)}
      \right)
    \right)
    =
    \infty
  \end{equation*}
  and
  \begin{equation*}
    \lim_{N \rightarrow \infty} \frac{1}{\epsilon_N \log(N)} \left(
      \min_{t \in \R} \left(
        \sum_{k=-N}^{N} f_\sigma(k) \frac{1-\cos(\pi(t-k))}{\pi(t-k)}
      \right)
    \right)
    =
    -\infty .
  \end{equation*}
\end{theorem}
Theorem~\ref{th:divergence_speed_conj_shannon_oversampl} shows that in the case of
oversampling, we have the same divergence behavior and speed that was observed in
Theorem~\ref{th:divergence_speed_conj_shannon_no_oversampl}, i.e, the case without
oversampling.
That is, if we use oversampling as in
Theorem~\ref{th:divergence_speed_conj_shannon_oversampl}, we have no improvement. 
Of course, due to oversampling, we have the freedom to use better, faster decaying kernels
than those in Theorem~\ref{th:divergence_speed_conj_shannon_oversampl}. 
We will analyze this situation in Section~\ref{sec:overs-with-kernels}.

\begin{IEEEproof}
  Let $\sigma \in(0,\pi]$ be arbitrary but fixed. 
  Further, let $\{\epsilon_N\}_{N \in \N}$ be an arbitrary sequence of positive numbers
  converging to zero, and $\bar{\epsilon}_N = \max_{M \geq N} \epsilon_M$, $N \in \N$. 
  Let $\{N_k\}_{k \in \N}$ be a strictly monotonically increasing sequence of natural
  numbers, such that $\bar{\epsilon}_{N_k} > \bar{\epsilon}_{N_{k+1}}$, $k \in \N$. 
  We set $\delta_k = \sqrt{\bar{\epsilon}_{N_k}} - \sqrt{\bar{\epsilon}_{N_{k+1}}}$,
  $k \in \N$. 
  For the proof we use the function $f_1$ from
  Theorem~\ref{th:divergence_speed_conj_shannon_no_oversampl}, which is defined in
  \eqref{eq:pr:th:divergence_speed_conj_shannon_no_oversampl:def_f1_wk}. 
  Let
  \begin{equation*}
    \Ft{f}_{\sigma} (\omega)
    =
    \begin{cases}
      \Ft{f}_1(\omega), &\lvert \omega \rvert < \sigma, \\
      0, & \sigma \leq \lvert \omega \rvert \leq \pi 
    \end{cases}
  \end{equation*}
  and
  \begin{equation*}
    \Ft{r}_{\sigma} (\omega)
    =
    \begin{cases}
      0, &\lvert \omega \rvert < \sigma, \\
      \Ft{f}_1(\omega), &\sigma \leq \lvert \omega \rvert \leq \pi .
    \end{cases}
  \end{equation*}
  Since
  \begin{equation*}
    \Ft{w}_{N_{k+1}}(\omega)
    =
    2 K_{2N_{k+1}}^{\text{F}}(\omega) - K_{N_{k+1}}^{\text{F}}(\omega) ,
  \end{equation*}
  where $K_N^{\text{F}}(\omega)$ denotes the Fej{\'e}r kernel
  \begin{equation*}
    K_N^{\text{F}}(\omega)
    =
    \frac{1}{N} \frac{\sin^2 \left( \frac{N \omega}{2} \right)}{\sin^2 \left(
          \frac{\omega}{2} \right)} ,
  \end{equation*}
  we see that, for $\omega \in [-\pi,-\sigma] \cup [\sigma,\pi]$, we have
  \begin{equation*}
    \lvert \delta_k \Ft{w}_{N_{k+1}}(\omega) \rvert
    \leq
    \frac{3 \delta_k}{N_{k+1} \sin^2\left( \frac{\sigma}{2} \right)} .
  \end{equation*}
  Further, since
  \begin{equation*}
    \sum_{k=1}^{\infty} \frac{3 \delta_k}{N_{k+1} \sin^2\left( \frac{\sigma}{2}
      \right)}
    <
    \infty ,
  \end{equation*}
  it follows that
  \begin{equation*}
    \sum_{k=1}^{\infty} \delta_k \Ft{w}_{N_{k+1}}
  \end{equation*}
  converges uniformly on $[-\pi,-\sigma] \cup [\sigma,\pi]$, and hence defines a
  continuous limit function $\Ft{g}$ on $[-\pi,-\sigma] \cup [\sigma,\pi]$. 
  It follows that $\Ft{g} \in L^2([-\pi,-\sigma] \cup [\sigma,\pi])$. 
  We already know from the proof of
  Theorem~\ref{th:divergence_speed_conj_shannon_no_oversampl} that
  \begin{equation}\label{eq:L1_convergence_sum}
    \lim_{N \rightarrow \infty} \int_{-\pi}^{\pi} \left| \Ft{f}_1(\omega) - \sum_{k=1}^{N} \delta_k
      \Ft{w}_{N_{k+1}}(\omega) \right| \di{\omega}
    =
    0 .
  \end{equation}
  Thus, we have
  \begin{align*}
    \int_{\sigma \leq \lvert \omega \rvert \leq \pi} \lvert \Ft{f}_1(\omega) - \Ft{g}(\omega) \rvert \di{\omega}
    &=
    \int_{\sigma \leq \lvert \omega \rvert \leq \pi} \left| \Ft{f}_1(\omega) - \lim_{N
        \rightarrow \infty} \sum_{k=1}^{N} \delta_k
      \Ft{w}_{N_{k+1}}(\omega) \right| \di{\omega} \\
    &=
    \lim_{N \rightarrow \infty} \int_{\sigma \leq \lvert \omega \rvert \leq \pi} \left|
      \Ft{f}_1(\omega) - \sum_{k=1}^{N} \delta_k \Ft{w}_{N_{k+1}}(\omega) \right|
    \di{\omega} \\
    &=
    0 ,
  \end{align*}
  where we used Lebesgue's dominated convergence theorem in the second to last and
  \eqref{eq:L1_convergence_sum} in the last equality. 
  This shows that $\Ft{f}_1 = \Ft{g}$ almost everywhere on
  $[-\pi,-\sigma] \cup [\sigma,\pi]$. 
  Hence, using the definition of $\Ft{r}_\sigma$, we see that
  $\Ft{r}_\sigma = \Ft{f}_1 = \Ft{g}$ almost everywhere on
  $[-\pi,-\sigma] \cup [\sigma,\pi]$. 
  Since $\Ft{r}_\sigma(\omega) = 0$ for all $\omega \in (-\sigma,\sigma)$, it follows that
  $\Ft{r}_\sigma \in L^2[-\pi,\pi]$, which in turn implies that
  $r_\sigma \in \PWs_{\pi}^{2}$. 
  Knowing that $r_\sigma \in \PWs_{\pi}^{2}$, it follows that
  \begin{equation*}
    \left|
      \sum_{k=-N}^{N} r_{\sigma}(k) \frac{1-\cos(\pi(t-k))}{\pi(t-k)}
    \right|
    \leq
    \lVert r_{\sigma} \rVert_{\PWs_{\pi}^{2}} 
  \end{equation*}
  for all $N \in \N$ and $t \in \R$, which in turn implies
  \begin{align*}
    \left|
      \sum_{k=-N}^{N} f_1(k) \frac{1-\cos(\pi(t-k))}{\pi(t-k)}
      -
      \sum_{k=-N}^{N} f_{\sigma}(k) \frac{1-\cos(\pi(t-k))}{\pi(t-k)}
    \right|
    &\leq
    \lVert r_{\sigma} \rVert_{\PWs_{\pi}^{2}} 
  \end{align*}
  for all $N \in \N$ and $t \in \R$.
  It follows that
  \begin{align*}
    &\frac{1}{\epsilon_N \log(N)} \left( \max_{t\in \R}
    \left( 
      \sum_{k=-N}^{N} f_{\sigma}(k) \frac{1-\cos(\pi(t-k))}{\pi(t-k)}
    \right) \right) \\
    &\geq
    \frac{1}{\epsilon_N \log(N)}
    \left(
      \max_{t\in \R} \left(
        \sum_{k=-N}^{N} f_1(k) \frac{1-\cos(\pi(t-k))}{\pi(t-k)}
      \right)
      -
      \lVert r_{\sigma} \rVert_{\PWs_{\pi}^{2}}
    \right)
  \end{align*}
  as well as
    \begin{align*}
    &\frac{1}{\epsilon_N \log(N)} \left( \min_{t\in \R}
    \left( 
      \sum_{k=-N}^{N} f_{\sigma}(k) \frac{1-\cos(\pi(t-k))}{\pi(t-k)}
    \right) \right) \\
    &\leq
    \frac{1}{\epsilon_N \log(N)}
    \left(
      \min_{t\in \R} \left(
        \sum_{k=-N}^{N} f_1(k) \frac{1-\cos(\pi(t-k))}{\pi(t-k)}
      \right)
      +
      \lVert r_{\sigma} \rVert_{\PWs_{\pi}^{2}}
    \right),
  \end{align*}
  which, together with Theorem~\ref{th:divergence_speed_conj_shannon_no_oversampl},
  completes the proof.
\end{IEEEproof}

Next, we come to the Shannon sampling series for the case of critical sampling at Nyquist
rate. 
In \cite{boche14c} it has been proved that there exists a function $f \in \PWs_{\pi}^{1}$
such that $\lVert S_N f \rVert_{\infty}$ diverges strongly, i.e., that
$\lim_{N \rightarrow \infty} \lVert S_N f \rVert_{\infty} = \infty$, and thus shown that
the answer to Question Q1 is negative. 
However, in \cite{boche14c} the authors also raised a question regarding the divergence
order. 
Using the function $f_1$ from the proof of
Theorem~\ref{th:divergence_speed_conj_shannon_no_oversampl}, it is possible to answer this
question.

\begin{theorem}\label{th:divergence_speed_shannon_no_oversampl}
  Let $\{\epsilon_N\}_{N \in \N}$ be an arbitrary sequence of positive numbers converging to zero.
  There exists a function $f_2 \in \PWs_{\pi}^{1}$ such that
  \begin{equation*}
    \lim_{N \rightarrow \infty} \frac{1}{\epsilon_N \log(N)} \left(
      \max_{t \in \R} \left(
        \sum_{k=-N}^{N} f_2(k) \frac{\sin(\pi(t-k))}{\pi(t-k)}
      \right)
    \right)
    =
    \infty
  \end{equation*}
  and
  \begin{equation*}
    \lim_{N \rightarrow \infty} \frac{1}{\epsilon_N \log(N)} \left(
      \min_{t \in \R} \left(
        \sum_{k=-N}^{N} f_2(k) \frac{\sin(\pi(t-k))}{\pi(t-k)}
      \right)
    \right)
    =
    -\infty .
  \end{equation*}
\end{theorem}

Theorem~\ref{th:divergence_speed_shannon_no_oversampl} shows that for the Shannon sampling
series it is possible to have strong divergence with order $\epsilon_N \log(N)$ for all
zero sequences $\epsilon_N$.

\begin{IEEEproof}
  Let $\{\epsilon_N\}_{N \in \N}$ be an arbitrary sequence of positive numbers converging
  to zero, and $\bar{\epsilon}_N = \max_{M \geq N} \epsilon_M$, $N \in \N$. 
  Let $\{N_k\}_{k \in \N}$ be a strictly monotonically increasing sequence of natural
  numbers, such that $\bar{\epsilon}_{N_k} > \bar{\epsilon}_{N_{k+1}}$, $k \in \N$. 
  We set $\delta_k = \sqrt{\bar{\epsilon}_{N_k}} - \sqrt{\bar{\epsilon}_{N_{k+1}}}$,
  $k \in \N$. 
  For the proof we use the function $f_1$ from
  Theorem~\ref{th:divergence_speed_conj_shannon_no_oversampl}, which is defined in
  \eqref{eq:pr:th:divergence_speed_conj_shannon_no_oversampl:def_f1_wk}. 
  Let $F_1(\e^{\iu \omega}) = f_1(\omega)$, $\omega \in [-\pi,\pi)$, and
  $F_2(\e^{\iu \omega}) = F_1(\e^{\iu (\omega+\pi)})$, $\omega \in \R$. 
  We have $F_1 \in L^1(\partial D)$ and consequently $F_2 \in L^1(\partial D)$, where
  $L^1(\partial D)$ denotes the set of Lebesgue measurable functions $F$ on the unit
  circle satisfying
  \begin{equation*}
    \frac{1}{2\pi} \int_{-\pi}^{\pi} \lvert F(\e^{\iu \omega}) \rvert \di{\omega}
    <
    \infty .
  \end{equation*}
  Further, let
  \begin{equation*}
    f_2(t)
    =
    \frac{1}{2\pi} \int_{-\pi}^{\pi} F_2(\e^{\iu \omega}) \e^{\iu \omega t} \di{\omega} .
  \end{equation*}
  It follows that $f_2 \in \PWs_{\pi}^{1}$,
  $\lVert f_2 \rVert_{\PWs_{\pi}^{1}} = \lVert f_1 \rVert_{\PWs_{\pi}^{1}} < \infty$, and
  \begin{align*}
    f_2(k)
    &=
    \frac{1}{2\pi} \int_{-\pi}^{\pi} F_2(\e^{\iu \omega}) \e^{\iu \omega k} \di{\omega} \\
    &=
    \frac{1}{2\pi} \int_{-\pi}^{\pi} F_1(\e^{\iu (\omega + \pi)}) \e^{\iu \omega k} \di{\omega} \\
    &=
    (-1)^k \frac{1}{2\pi} \int_{0}^{2 \pi} F_1(\e^{\iu \xi}) \e^{\iu \xi k} \di{\xi} \\
    &=
    (-1)^k f_1(k) .
  \end{align*} 
  For $N \in \N$, $N$ even, and $t_N = N + 1/2$ we have
  \begin{align}
    \sum_{k=-N}^{N} f_2(k) \frac{\sin(\pi(t_N-k))}{\pi(t_N-k)}
    &=
    \sum_{k=-N}^{N} f_2(k) \frac{\sin(\pi(N+\frac{1}{2}-k))}{\pi(N+\frac{1}{2}-k)}
    \notag \\
    &=
    \frac{1}{\pi} \sum_{k=-N}^{N} (-1)^k f_1(k) \frac{(-1)^k}{\pi(N+\frac{1}{2}-k)}
    \notag \\
    &=
    \frac{1}{\pi} \sum_{k=-N}^{N} f_1(k) \frac{1}{N+\frac{1}{2}-k}
    \notag \\
    &\geq
    \frac{1}{\pi} \sum_{k=-N}^{N} f_1(k) \frac{1}{N+\frac{3}{2}-k} , \label{eq:shannon_div_speed_N_even}
  \end{align}
  because $f_1(k) \geq 0$ for all $k \in \Z$.
  For $N \in \N$, $N$ odd, and $t_N = N + 3/2$ we have
  \begin{align}
    \sum_{k=-N}^{N} f_2(k) \frac{\sin(\pi(t_N-k))}{\pi(t_N-k)}
    &=
    \sum_{k=-N}^{N} f_2(k) \frac{\sin(\pi(N+\frac{3}{2}-k))}{\pi(N+\frac{3}{2}-k)}
    \notag \\
    &=
    \frac{1}{\pi} \sum_{k=-N}^{N} (-1)^k f_1(k) \frac{(-1)^k}{\pi(N+\frac{3}{2}-k)}
    \notag \\
    &=
    \frac{1}{\pi} \sum_{k=-N}^{N} f_1(k) \frac{1}{N+\frac{3}{2}-k} . \label{eq:shannon_div_speed_N_odd}
  \end{align}
  Hence, we see from \eqref{eq:shannon_div_speed_N_even} and
  \eqref{eq:shannon_div_speed_N_odd} that
  \begin{equation}\label{eq:max_sum_geq_sum}
    \max_{t \in \R} \left( \sum_{k=-N}^{N} f_2(k) \frac{\sin(\pi(t-k))}{\pi(t-k)} \right)
    \geq
    \frac{1}{\pi} \sum_{k=-N}^{N} f_1(k) \frac{1}{N+\frac{3}{2}-k}
  \end{equation}
  for all $N \in \N$.
  Using the same calculation as in the proof of
  Theorem~\ref{th:divergence_speed_conj_shannon_no_oversampl}, it is shown that
  \begin{equation}\label{eq:shannon_div_speed_2}
    \frac{1}{\pi} \sum_{k=-N}^{N} f_1(k) \frac{1}{N+\frac{3}{2}-k}
    \geq
    \frac{1}{\pi} \epsilon_N \log(N) \frac{1}{\sqrt{\epsilon_N}} .
  \end{equation}
  From \eqref{eq:max_sum_geq_sum} and \eqref{eq:shannon_div_speed_2} it follows that
  \begin{align*}
    \lim_{N \rightarrow \infty} \frac{1}{\epsilon_N \log(N)} \left(
      \max_{t \in \R} \left(
        \sum_{k=-N}^{N} f_2(k) \frac{\sin(\pi(t-k))}{\pi(t-k)}
      \right)
    \right)
    =
    \infty ,
  \end{align*}
  which proves the first assertion of the theorem.
  The second assertion is proved similarly by choosing $t_N = N + 3/2$ instead of
  $t_N = N + 1/2$ in \eqref{eq:shannon_div_speed_N_even} and $t_N = N + 1/2$ instead of
  $t_N = N + 3/2$ in \eqref{eq:shannon_div_speed_N_odd}.
\end{IEEEproof}

In the next section we analyze the use of more general kernels.

\section{Oversampling with Kernels}\label{sec:overs-with-kernels}

We now come back to the situation where we know the function $f$ on an oversampling set. 
In Theorem~\ref{th:divergence_speed_conj_shannon_oversampl} we already studied the
oversampling case for the conjugated Shannon sampling series and observed that mere
oversampling with the standard kernel does not remove the divergence. 
However, the redundance introduced by oversampling allows us to use other, faster decaying
kernels. 
This introduces a further degree of freedom that can be employed for adaptivity. 
In addition to the subsequence $\{N_n\}_{n \in \N}$, we now can also choose the
reconstruction kernel dependently on the signal $f$. 
Thus, question Q1 can be extended in the case of oversampling to also include the adaptive
choice of the kernel. 
We will show in this section that for any amount of oversampling the extended
question Q1 has to be answered negatively. 
That is, even the joint optimization of the choice of the subsequence $\{N_n\}_{n \in \N}$
and the reconstruction kernel cannot circumvent the divergence.

We first consider the function reconstruction problem. 
In the oversampling case, it is possible to create absolutely convergent sampling series
by using other kernels than the $\sinc$-kernel \cite{butzer86,engels87,butzer88}. 
In particular, all kernels $\phi$ in the set $\mathcal{M}(a)$, which is defined next, can
be used.
\begin{definition}
  $\mathcal{M}(a)$, $a > 1$, is the set of functions $\phi \in \Bs_{a \pi}^{1}$ with
  $\Ft{\phi}(\omega)=1/a$ for $|\omega| \leq \pi$.
\end{definition}
The functions in $\mathcal{M}(a)$, $a>1$, are suitable kernels for the sampling series,
because for all $ f\in \PWs_{\pi}^{1}$ and $a>1$ we have
\begin{equation*}
  \lim_{N \rightarrow \infty} \max_{t \in \R} \left|  f(t)- \sum_{k=-N}^{N} f \left(\frac{k}{a}\right)
      \phi\left(t -\frac{k}{a}\right) \right| =0
\end{equation*}
if $\phi \in \mathcal{M}(a)$.

We introduce the abbreviation
\begin{equation*}
  (H_{N,\phi}^{a} f)(t)
  \defequal
  \sum_{k=-N}^{N} f \left( \frac{k}{a} \right) (\Hto \phi) \left( t-\frac{k}{a}
  \right) .
\end{equation*}

\begin{theorem}\label{th:hilbert_oversampling_global_strong_plus_minus}
  Let $\{\epsilon_N\}_{N \in \N}$ be an arbitrary sequence of positive numbers converging
  to zero.
  There exists a universal function $f_1 \in \PWs_{\pi}^{1}$ such that for all $a>1$ and
  for all $\phi \in \mathcal{M}(a)$ we have
  \begin{equation*}
    \lim_{N \rightarrow \infty} \frac{1}{\epsilon_N \log(N)} \max_{t \in \R}
    (H_{N,\phi}^{a} f_1)(t)
    =
    \infty
  \end{equation*}
  and
  \begin{equation*}
    \lim_{N \rightarrow \infty} \frac{1}{\epsilon_N \log(N)} \min_{t \in \R}
    (H_{N,\phi}^{a} f_1)(t)
    =
    -\infty .
  \end{equation*} 
\end{theorem}
Theorem~\ref{th:hilbert_oversampling_global_strong_plus_minus} shows that it is possible
to have strong divergence with order $\epsilon_N \log(N)$ for all zero sequences
$\epsilon_N$ even in the case of oversampling.
\begin{remark}\label{re:log_N}
  We have the following result. 
  Let $a>1$ be arbitrary. 
  For every $\phi \in \mathcal{M}(a)$ there exists a constant $C_{1}$ such
  that
\begin{equation*}
  \lVert H_{N,\phi}^{a} f \rVert_{\infty}
  \leq
  C_{1} \log(N) \lVert f \rVert_{\PWs_{\pi}^{1}}
\end{equation*}
for all $N \geq 2$ and all $f \in \PWs_{\pi}^{1}$.
It follows that
\begin{equation*}
  \lim_{N \rightarrow \infty}
  \frac{\lVert H_{N,\phi}^{a} f \rVert_{\infty}}{\log(N)}
  =
  0 .
\end{equation*}
This shows how sharp the result in
Theorem~\ref{th:hilbert_oversampling_global_strong_plus_minus} is. 
Note that the same result is also true for Theorems
\ref{th:divergence_speed_conj_shannon_no_oversampl}--\ref{th:divergence_speed_shannon_no_oversampl}.  
\end{remark}

\begin{remark}
  As already mentioned, Erd{\H{o}}s analyzed the question of strong divergence for the
  Lagrange interpolation on Chebyshev nodes in \cite{erdos41}. 
  There, for continuous functions, a similar $\log(N)$ upper bound, as in
  Remark~\ref{re:log_N}, holds for the maximum norm of the Lagrange interpolation
  polynomials. 
  The original problem that was formulated in \cite{erdos41} is still open.
  However, an analysis of the behavior of Lagrange interpolation polynomials indicates
  that even if strong divergences occurs, a statement like in
  Theorem~\ref{th:divergence_speed_shannon_no_oversampl} about the $\epsilon_N \log(N)$
  divergence speed cannot hold, i.e., it is not possible to get arbitrarily ``close'' to
  $\log(N)$ divergence. 
\end{remark}

The proof of Theorem~\ref{th:hilbert_oversampling_global_strong_plus_minus} uses some
techniques and the following lemma from \cite{boche10b}.
\begin{lemma}\label{le:upper_bound_sum_r}
  For all $a>1$, $f \in \PWs_{\pi}^{1}$, $N \in \N$ and $\lvert t \rvert \geq (N+1)/a$ we have
  \begin{equation*}
    \sum_{k=-N}^N \left| f
      \left( \frac{k}{a} \right) r \left( t-\frac{k}{a} \right)\right|
    <
    a^2 \lVert f \rVert_\infty ,
  \end{equation*}
  where
  \begin{equation*}
    r(t)
    \defequal
    \frac{2}{\pi^2 t^2} \left( \sin(\pi t) -
      \sin\left(\frac{\pi}{2} t \right)  \right) .
  \end{equation*}
\end{lemma}

\begin{IEEEproof}[Proof of Theorem~\ref{th:hilbert_oversampling_global_strong_plus_minus}]
\begin{figure}
\centering
\begin{tikzpicture}[xscale=0.9,yscale=0.9]
  \footnotesize
  \draw[->,>=latex] (-5,0)--(5.1,0) node[below] {$\omega$};
  \draw[->,>=latex] (0,-0.2)--(0,1.9);
      
  \draw node[above right] at (0,1) {$1$};

  \draw node[below] at (1,-0.1) {$\frac{\pi}{2}$}; \draw (1,0.1) --
  (1,-0.1);

  \draw node[below] at (-1,-0.1) {$-\frac{\pi}{2}$}; \draw (-1,0.1) --
  (-1,-0.1);

  \draw node[below] at (2,-0.1) {$\pi$}; \draw (2,0.1) --
  (2,-0.1);

  \draw node[below] at (-2,-0.1) {$-\pi$}; \draw (-2,0.1) --
  (-2,-0.1);

  \draw[semithick] (-5,0)--(-2,0)--(-1,1)--(1,1)--(2,0)--(5,0);
  \draw[semithick,dashed]
  (-5,0)--(-4.5,0)--(-3.5,1)--(-2,1)--(-1,0)--(1,0)--(2,1)--(3.5,1)--(4.5,0)--(5,0);
  
  \draw node[below] at (3.5,-0.1) {$a\pi$}; \draw (3.5,0.1) --
  (3.5,-0.1);

  \draw node[below] at (-3.5,-0.1) {$-a\pi$}; \draw (-3.5,0.1) --
  (-3.5,-0.1);

  \draw node[above] at (-0.6,1) {$\Ft{q}_1$};
  \draw node[above] at (2.9,1) {$\Ft{q}_2$};
\end{tikzpicture}
\caption{Definition of $\Ft{q}_1$ (solid line) and $\Ft{q}_2$ (dashed line).}
\label{fig:definition_q}
\end{figure}
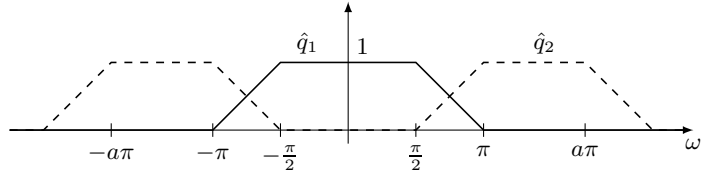
Let $a>1$ be arbitrary but fixed. 
Furthermore, let $\Ft{q}_1$ and $\Ft{q}_2$ be the functions defined in Figure
\ref{fig:definition_q} and $\phi \in \mathcal{M}(a)$ some arbitrary reconstruction kernel. 
Then we have
\begin{align*}
  \phi
  =
  \phi \convolution q_1 + \phi \convolution q_2 
  =
  q_1 + \phi \convolution q_2
\end{align*}
and 
\begin{equation*}
  \Hto \phi
  =
  \Hto q_1 + \Hto (\phi \convolution q_2)
  =
  \Hto q_1 + \phi \convolution (\Hto q_2) .  
\end{equation*}
Since $\Hto q_2 \in L^1(\R)$, it follows that
$s \defequal \phi \convolution (\Hto q_2) \in L^1(\R)$.
Moreover, for $N \in \N$ and $f \in \PWs_{\pi}^{1}$ we have
  \begin{align}
    &\left| (H_{N,\phi}^{a} f)(t) - (\Hto_{N,q_1}^a f)(t) \right| \notag \\
    &\qquad=
    \left|
      \sum_{k=-N}^{N}
      f \left( \frac{k}{a} \right) (\Hto \phi)\left( t-\frac{k}{a}
      \right)
      -
      \sum_{k=-N}^{N}
      f \left( \frac{k}{a} \right) (\Hto q_1)\left( t-\frac{k}{a}
      \right)
    \right| \notag\\
    &\qquad=
    \left| 
      \sum_{k=-N}^{N}
      f \left( \frac{k}{a} \right) s\left( t-\frac{k}{a}
      \right)
    \right| \notag\\
    &\qquad\leq
    \sum_{k=-N}^{N}
    \left|  f \left( \frac{k}{a} \right)  \right|
    \left| s\left( t-\frac{k}{a} \right)  \right| \notag\\
    &\qquad\leq
    \lVert f \rVert_{\infty} \sum_{k=-\infty}^{\infty}
    \left| s\left(t-\frac{k}{a} \right) \right| \notag\\
    &\qquad\leq
    C_{2} \lVert f \rVert_{\infty} \lVert s
    \rVert_{\Bs_{a \pi}^1} , \label{eq:difference_leq}
  \end{align}
  where we used Nikol'ski\u{\i}'s inequality \cite[p.~49]{higgins96_book} in the last
  step.
  For $\tau \neq 0$ we can simplify $(\Hto q_1)(\tau)$, using integration by parts,
  according to
  \begin{align*}
  (\Hto q_1)(\tau)
  &=
  \frac{1}{2\pi} \int_{-\pi}^{\pi} - \iu \sgn(\omega) \Ft{q}_1(\omega)
  \e^{\iu \omega \tau} \di{\omega} \\
  &=
  \frac{1}{\pi} \int_{0}^{\pi} \sin(\omega \tau) \Ft{q}_1(\omega)
  \di{\omega} \\
  &=
  \frac{1}{\pi \tau} - r(\tau) ,
\end{align*}
where
\begin{equation*}
  r(\tau)
  \defequal
  \frac{2}{\pi^2 \tau^2} \left( \sin(\pi \tau) - \sin\left(\frac{\pi}{2} \tau \right)  \right) .
\end{equation*}
For $\lvert t \rvert \geq (N+1)/a$ we thus obtain
\begin{equation*}
  (\Hto_{N,q_1}^a f)(t)
  =
  \sum_{k=-N}^{N}
  f \left( \frac{k}{a} \right) \frac{1}{\pi \left( t-\frac{k}{a}
  \right)}
  -
  \sum_{k=-N}^{N}
  f \left( \frac{k}{a} \right) r\left( t-\frac{k}{a}
  \right) ,
\end{equation*}
and since 
\begin{equation*}
  \left|
    \sum_{k=-N}^{N}
    f \left( \frac{k}{a} \right) r\left( t-\frac{k}{a} \right)
  \right|
  <
  a^2 \lVert f \rVert_{\infty}
\end{equation*}
by Lemma~\ref{le:upper_bound_sum_r}, it follows that
\begin{equation}\label{eq:HNq1af_g_sum_minus_const}
  (\Hto_{N,q_1}^a f)(t)
  >
  \sum_{k=-N}^{N}
  f \left( \frac{k}{a} \right) \frac{1}{\pi \left( t-\frac{k}{a}
  \right)}
  -
  a^2 \lVert f \rVert_{\infty} .
\end{equation}
Combining \eqref{eq:difference_leq} and \eqref{eq:HNq1af_g_sum_minus_const} we see that
\begin{align}
  (H_{N,\phi}^{a} f)(t)
  &\geq
  (\Hto_{N,q_1}^a f)(t) - C_{2} \lVert f \rVert_{\infty} \lVert s
    \rVert_{\Bs_{a \pi}^1} \notag \\
  &>
  \sum_{k=-N}^{N}
  f \left( \frac{k}{a} \right) \frac{1}{\pi \left( t-\frac{k}{a}
  \right)}
  -  (a^2 + C_{2} \lVert s \rVert_{\Bs_{a \pi}^1}) \lVert f \rVert_{\infty} \label{eq:HNphiaf_g_term}
\end{align}
for all $\lvert t \rvert \geq (N+1)/a$ and all $f \in \PWs_{\pi}^{1}$. 
Hence, it suffices to concentrate the analysis on
\begin{equation*}
  \sum_{k=-N}^{N} f \left( \frac{k}{a} \right) \frac{1}{\pi \left( t-\frac{k}{a} \right)}
\end{equation*}
in the following.

Let $\{\epsilon_N\}_{N \in \N}$ be an arbitrary sequence of positive numbers converging to
zero, and $\bar{\epsilon}_N = \max_{M \geq N} \epsilon_M$, $N \in \N$. 
Note that $\bar{\epsilon}_N \geq \epsilon_N$ for all $N \in \N$. 
Further, let $\{N_k\}_{k \in \N}$ be a strictly monotonically increasing sequence of
natural numbers, such that $\bar{\epsilon}_{N_k} > \bar{\epsilon}_{N_{k+1}}$, $k \in \N$. 
We set $\delta_k = \sqrt{\bar{\epsilon}_{N_k}} - \sqrt{\bar{\epsilon}_{N_{k+1}}}$,
$k \in \N$. 
It follows that $\delta_k > 0$ for all $k \in \N$ and
\begin{equation}\label{eq:sum_delta_k}
  \sum_{k=1}^{\infty} \delta_k = \sqrt{\bar{\epsilon}_{N_1}} < \infty .
\end{equation}
For $M \in \N$ we consider the functions
\begin{equation*}
  g_M(t)
  =
  \left( \frac{\sin \left( \frac{\pi}{M} t \right)}{\frac{\pi}{M} t} \right)^2 .
\end{equation*}
Note that $\lVert g_M \rVert_{\PWs_{\pi}^{1}} = 1$ for all $M \in \N$. 
Let $\{M_k\}_{k \in \N}$ be a sequence of monotonically increasing natural numbers, such
that $g_{M_k}(t) \geq 1/2$ for $\lvert t \rvert \leq N_{k+1}$, $k \in \N$. 
We define the function
\begin{equation}\label{eq:def_f1}
  f_1
  =
  \sum_{k=1}^{\infty} \delta_{k} g_{M_k} .
\end{equation}
Since $\lVert \delta_{k} g_{M_k} \rVert_{\PWs_{\pi}^{1}} = \delta_{k}$ and because of
\eqref{eq:sum_delta_k}, it follows that the series in \eqref{eq:def_f1} converges in the
$\PWs_{\pi}^{1}$-norm and consequently uniformly on $\R$.

Let $N \in \N$ be arbitrary but fixed. 
There exists exactly one $\hat{k} \in \N$ such that $N \in [N_{\hat{k}}, N_{\hat{k}+1})$. 
Since $\delta_k > 0$ and $g_{M_k}(t) \geq 0$ for all $t \in \R$ and all $k \in \N$, we
have, for $t_N^{(1)} = (N+1)/a$, that
\begin{align*}
  \sum_{l=-N}^{N} f_1 \left( \frac{l}{a} \right) \frac{1}{\pi \left( t_N^{(1)} -\frac{l}{a}
    \right)}
  &\geq
  \sum_{l=-N}^{N}  \sum_{k=\hat{k}}^{\infty} \delta_{k} g_{M_k}
    \left( \frac{l}{a} \right) \frac{1}{\pi \left(t_N^{(1)}-\frac{l}{a} \right)} \\
  &\geq
  \frac{1}{2} \sum_{k=\hat{k}}^{\infty} \delta_{k} \sum_{l=-N}^{N}
  \frac{1}{\pi \left( t_N^{(1)}-\frac{l}{a} \right)} ,
\end{align*}
where we used in the second inequality that
\begin{equation*}
  g_{M_k} \left( \frac{l}{a} \right)
  \geq
  \frac{1}{2}
\end{equation*}
for all $k \geq \hat{k}$ and all $\lvert l \rvert \leq N$.
It follows that
\begin{align}
  \sum_{l=-N}^{N} f_1 \left( \frac{l}{a} \right) \frac{1}{\pi \left( t_{N}^{(1)}-\frac{l}{a}
    \right)}
  &\geq
  \frac{1}{2} \sum_{k=\hat{k}}^{\infty} \delta_{k} \sum_{l=-N}^{N}
  \frac{1}{\pi \left( \frac{N+1}{a}-\frac{l}{a} \right)} \notag\\
  &=
  \frac{a}{2 \pi} \sum_{k=\hat{k}}^{\infty} \delta_{k} \sum_{l=1}^{2N+1}
  \frac{1}{l} \notag\\
  &\geq
  \frac{a}{2 \pi} \log(2N+2) \sum_{k=\hat{k}}^{\infty} \delta_{k} \notag\\
  &=
  \frac{a}{2 \pi} \log(2N+2) \sqrt{\bar{\epsilon}_{N_{\hat{k}}}} \notag\\
  &\geq
  \frac{a}{2 \pi} \epsilon_{N} \log(N)
  \frac{1}{\sqrt{\epsilon_{N}}} , \label{eq:sum_geq_epsn_logn}
\end{align}
because $N \geq N_{\hat{k}}$ and thus
$\sqrt{\bar{\epsilon}_{N_{\hat{k}}}}
  \geq
  \sqrt{\epsilon_{N_{\hat{k}}}}
  \geq
  \sqrt{\epsilon_{N}}$.
From \eqref{eq:sum_geq_epsn_logn} we see that
\begin{equation*}
  \lim_{N \rightarrow \infty} \frac{1}{\epsilon_N \log(N)}
    \sum_{l=-N}^{N} f_1 \left( \frac{l}{a} \right)
    \frac{1}{\pi \left( t_{N}^{(1)}-\frac{l}{a} \right)} 
  =
  \infty .
\end{equation*}
Thus, it follows from \eqref{eq:HNphiaf_g_term} that, for arbitrary $a>1$ and
$\phi \in \mathcal{M}(a)$, we have
\begin{equation*}
  \lim_{N \rightarrow \infty} \frac{1}{\epsilon_N \log(N)} \max_{t \in \R}
  (H_{N,\phi}^{a} f_1)(t)
  =
  \infty .
\end{equation*}

Following the same line of reasoning it is shown that, for
$t_{N}^{(2)} = -(N+1)/a$, we have
\begin{equation*}
  \lim_{N \rightarrow \infty} \frac{1}{\epsilon_N \log(N)}
    \sum_{l=-N}^{N} f_1 \left( \frac{l}{a} \right)
    \frac{1}{\pi \left( t_{N}^{(2)}-\frac{l}{a} \right)}
  =
  -\infty ,
\end{equation*}
and consequently
\begin{equation*}
  \lim_{N \rightarrow \infty} \frac{1}{\epsilon_N \log(N)} \min_{t \in \R}
  (H_{N,\phi}^{a} f_1)(t)
  =
  -\infty .
\end{equation*}
\end{IEEEproof}

As explained in the introduction, it is interesting and also important for applications to
analyze the general question when and why strong divergence occurs. 
We have already seen several cases in this paper where strong divergence emerged, however
a general theory is missing.

\section{Pointwise Convergence Behavior}

In Sections~\ref{sec:behav-conj-shann} and \ref{sec:overs-with-kernels} we analyzed the
global behavior of the reconstruction and approximation processes. 
In this section we will study the pointwise behavior of the system approximation process
for fixed $t \in \R$, i.e., the quantity of interest is $(T_N f)(t)$. 
We want to know if Question Q1 has to be answered negatively in this case. 
It will turn out that the situation is different, and that Question Q1 has a positive
answer for all stable LTI systems $T$ and all $t \in \R$.

Let $T_{N,t} f \defequal (T_N f)(t)$. For $t \in \R$ we consider
\begin{equation*}
  \lVert T_{N,t} \rVert_*
  \defequal
  \sup_{\lVert f \rVert_{\PWs_{\pi}^{1}} \leq 1} \lvert (T_N f)(t) \rvert .
\end{equation*}
It is known that for every $t \in \R$ there exists a stable LTI system
$T^1 \colon \PWs_{\pi}^{1} \to \PWs_{\pi}^{1}$ such that
\begin{equation*}
  \limsup_{N \rightarrow \infty} \lVert T_{N,t}^1 \rVert_*
  =
  \infty .
\end{equation*}
Therefore, there exists a function $f_1 \in \PWs_{\pi}^{1}$ such that
\begin{equation}\label{eq:limsup_TNf_fixed_t}
  \limsup_{N \rightarrow \infty} \lvert (T_N^1 f_1)(t) \rvert
  =
  \infty.
\end{equation}
This shows that for every $t \in \R$ there exists a stable LTI system
$T^1 \colon \PWs_{\pi}^{1} \to \PWs_{\pi}^{1}$ and a function $f_1 \in \PWs_{\pi}^{1}$
such that the system approximation process $(T_N^1 f_1)(t)$ diverges weakly.

Note that \eqref{eq:limsup_TNf_fixed_t} is true not only for equidistant sampling as in
\eqref{eq:def_TN}, but also for any sampling pattern that is a complete interpolating
sequence \cite{boche14d}.

The question whether $(T_N f)(t)$ also converges strongly for some stable LTI system $T$
and function $f \in \PWs_{\pi}^{1}$ is the topic of this section. 
It will turn out that strong divergence cannot occur in this case. 
Thus weak divergence does not automatically imply strong divergence. 
Hence, for the approximation process $(T_N f)(t)$, we can answer Question Q1 positively.

We first make a statement about the convergence of the Ces{\`a}ro  means
\begin{equation}\label{eq:arithmetic_mean_TNf}
  \frac{1}{M} \sum_{N=0}^{M-1} (T_Nf)(t) .
\end{equation}
The following theorem shows that \eqref{eq:arithmetic_mean_TNf} converges globally
uniformly, and consequently for fixed $t \in \R$, to $(Tf)(t)$ as $M$ tends to infinity.

\begin{theorem}\label{th:uniform_convergence_arith_mean}
  Let $T \colon \PWs_{\pi}^{1} \to \PWs_{\pi}^{1}$ be a stable LTI system. For all $f \in
  \PWs_{\pi}^{1}$ we have
  \begin{equation*}
    \lim_{M \rightarrow \infty} \max_{t \in \R} \left| (Tf)(t) - \frac{1}{M}
      \sum_{N=0}^{M-1} (T_Nf)(t) \right|
    =
    0 .
  \end{equation*}
\end{theorem}

\begin{IEEEproof}
  Let $T \colon \PWs_{\pi}^{1} \to \PWs_{\pi}^{1}$ be a stable LTI system, arbitrary but
  fixed. 
  For $f \in \PWs_{\pi}^{1}$, $t \in \R$, and $N \in \N_0 = \N \cup \{0\}$ we have
\begin{align*}
  (T_Nf)(t)
  &=
  \sum_{k=-N}^{N} f(k) h_T(t-k) \\
  &=
  \frac{1}{2\pi} \int_{-\pi}^{\pi} \Ft{f}(\omega) \frac{1}{2\pi} \int_{-\pi}^{\pi}
  \Ft{h}_T(\omega_1) \sum_{k=-N}^{N} \e^{\iu k(\omega-\omega_1)} \e^{\iu \omega_1 t}
  \di{\omega_1} \di{\omega}
\end{align*}
and it follows, for $M \in \N$, that
\begin{align}
  &\frac{1}{M} \sum_{N=0}^{M-1} (T_Nf)(t) \notag \\
  &\quad=
  \frac{1}{2\pi} \int_{-\pi}^{\pi} \Ft{f}(\omega) \frac{1}{2\pi} \int_{-\pi}^{\pi}
  \Ft{h}_T(\omega_1) \e^{\iu \omega_1 t} \frac{1}{M} \sum_{N=0}^{M-1}
  \left(
    \sum_{k=-N}^{N} \e^{\iu k(\omega-\omega_1)}
  \right) 
  \di{\omega_1} \di{\omega} \notag \\
  &\quad=
  \frac{1}{2\pi} \int_{-\pi}^{\pi} \Ft{f}(\omega) \frac{1}{2\pi} \int_{-\pi}^{\pi}
  \Ft{h}_T(\omega_1) \e^{\iu \omega_1 t}
  K_M^{\text{F}}(\omega - \omega_1)
  \di{\omega_1} \di{\omega} , \label{eq:arith_mean_TN}
\end{align} 
where
\begin{equation*}
  K_M^{\text{F}}
  =
  \frac{1}{M}
  \left( \frac{\sin \left( \frac{Mx}{2} \right)}{\sin \left( \frac{x}{2} \right)}
  \right)^2 , \qquad M \in \N,  
\end{equation*}
denotes the Fej\'er kernel.
We have $K_M^{\text{F}}(\omega) \geq 0$ for all $\omega \in [-\pi,\pi]$ and
\begin{equation*}
  \frac{1}{2\pi}
  \int_{-\pi}^{\pi} K_M^{\text{F}}(\omega) \di{\omega}
  =
  1 
\end{equation*}
for all $M \in \N$.
Since
\begin{align*}
  \left|
  \frac{1}{2\pi} \int_{-\pi}^{\pi}
  \Ft{h}_T(\omega_1) \e^{\iu \omega_1 t}
  K_M^{\text{F}}(\omega - \omega_1)
  \di{\omega_1}
  \right|
  &\leq
  \frac{1}{2\pi} \int_{-\pi}^{\pi}
  \lvert \Ft{h}_T(\omega_1) \rvert
  K_M^{\text{F}}(\omega - \omega_1)
  \di{\omega_1} \\
  &\leq
  \lVert T \rVert
  \frac{1}{2\pi} \int_{-\pi}^{\pi}
  K_M^{\text{F}}(\omega - \omega_1)
  \di{\omega_1} \\
  &=
  \lVert T \rVert
\end{align*}
for all $\omega \in [-\pi, \pi]$ and all $t\in \R$, we see from \eqref{eq:arith_mean_TN}
that
\begin{equation}\label{eq:av_arith_mean_TN_leq_norm_T_norm_f}
  \left| \frac{1}{M} \sum_{N=0}^{M-1} (T_Nf)(t) \right|
  \leq
  \lVert T \rVert \lVert f \rVert_{\PWs_{\pi}^{1}}
\end{equation}
for all $f \in \PWs_{\pi}^{1}$ and all $t \in \R$.

Let $f \in \PWs_{\pi}^{1}$ and $\epsilon \in (0,1)$ be arbitrary but fixed. 
There exists a $f_\epsilon \in \PWs_{\pi}^{2}$ such that
  \begin{equation*}
    \lVert f - f_\epsilon \rVert_{\PWs_{\pi}^{1}}
    \leq
    \epsilon .
  \end{equation*}
  Further, since $f_\epsilon \in \PWs_{\pi}^{2}$, there exists a natural number
  $N_0 = N_0(\epsilon)$, such that
  \begin{equation}\label{eq:max_t_Tf_eps_minus_TNf_eps_leq_eps}
    \max_{t \in \R} \left| (Tf_\epsilon)(t) - (T_N f_\epsilon)(t) \right|
    <
    \epsilon
  \end{equation}
  for all $N \geq N_0$.
  For $M \in \N$ we have
  \begin{align}
    &\left| (Tf)(t) - \frac{1}{M} \sum_{N=0}^{M-1} (T_N f)(t) \right| \notag \\
    &\quad=
    \Biggl| (Tf)(t) - (Tf_\epsilon)(t) + (Tf_\epsilon)(t) \notag \\
    &\qquad - \frac{1}{M} \sum_{N=0}^{M-1} (T_N f_\epsilon)(t)
      - \frac{1}{M} \sum_{N=0}^{M-1} (T_N (f-f_\epsilon))(t)
    \Biggr| \notag \\
    &\quad\leq
    \lvert (Tf)(t) - (Tf_\epsilon)(t) \rvert
    + \left| (Tf_\epsilon)(t) - \frac{1}{M} \sum_{N=0}^{M-1} (T_N f_\epsilon)(t)
    \right| \notag \\
    &\qquad + \left| \frac{1}{M} \sum_{N=0}^{M-1} (T_N (f-f_\epsilon))(t)
    \right| \notag \\
    &\quad\leq
    \lVert T \rVert \lVert f-f_\epsilon \rVert_{\PWs_{\pi}^{1}}
    +
    \left| \frac{1}{M} \sum_{N=0}^{M-1} (Tf_\epsilon)(t) - (T_N f_\epsilon)(t) \right| \notag \\
    &\qquad + \lVert T \rVert \lVert f-f_\epsilon \rVert_{\PWs_{\pi}^{1}} , \label{eq:Tf_minus_arithm_mean}
  \end{align}
  where we used \eqref{eq:av_arith_mean_TN_leq_norm_T_norm_f} in the last inequality. 
  For the second term on the right-hand side of \eqref{eq:Tf_minus_arithm_mean} we obtain,
  for $M \geq N_0+1$, that
  \begin{align}
    &\left| \frac{1}{M} \sum_{N=0}^{M-1} (Tf_\epsilon)(t) - (T_N f_\epsilon)(t) \right| \notag \\
    &\quad \leq
    \frac{1}{M} \sum_{N=0}^{N_0-1} \lvert (Tf_\epsilon)(t) - (T_N f_\epsilon)(t) \rvert
    +
    \frac{1}{M} \sum_{N=N_0}^{M-1} \lvert (Tf_\epsilon)(t) - (T_N f_\epsilon)(t) \rvert \notag \\
    &\quad \leq
    \frac{N_0}{M} \max_{0 \leq N \leq N_0-1} \lvert (Tf_\epsilon)(t) - (T_N f_\epsilon)(t)
    \rvert \notag \\
    &\qquad +
    \frac{M-N_0}{M} \max_{N_0 \leq N \leq M-1} \lvert (Tf_\epsilon)(t) - (T_N f_\epsilon)(t) \rvert \notag \\
    &\quad \leq
    \frac{N_0}{M} \max_{t \in \R} \max_{1 \leq N \leq N_0-1} \lvert (Tf_\epsilon)(t) - (T_N f_\epsilon)(t) \rvert
    +
    \frac{M-N_0}{M} \epsilon , \label{eq:second_term_conv_arith_mean}
  \end{align}
  where we used \eqref{eq:max_t_Tf_eps_minus_TNf_eps_leq_eps} in the last inequality. 
  We choose $M_0 \geq N_0+1$ large enough such that
  \begin{equation}\label{eq:M_zero_large_enough}
    \frac{N_0}{M_0} \max_{t \in \R} \max_{1 \leq N \leq N_0-1} \lvert (Tf_\epsilon)(t) -
    (T_N f_\epsilon))(t) \rvert
    <
    \epsilon .
  \end{equation}
  From \eqref{eq:Tf_minus_arithm_mean}, \eqref{eq:second_term_conv_arith_mean}, and
  \eqref{eq:M_zero_large_enough} it follows that
  \begin{equation}\label{eq:Tf_minus_arithm_mean_final}
    \left| (Tf)(t) - \frac{1}{M} \sum_{N=0}^{M-1} (T_N f)(t) \right|
    <
    2 \lVert T \rVert \epsilon + \epsilon
  \end{equation}
  for all $M \geq M_0 = M_0(\epsilon)$.
  Since the right-hand side of \eqref{eq:Tf_minus_arithm_mean_final} is independent of
  $t$, the proof is complete.
\end{IEEEproof}

Now we can answer the question from the beginning of this section whether we also have
strong divergence for fixed $t \in \R$ in the system approximation case.
\begin{theorem}\label{th:convergent_subsequence}
  Let $T \colon \PWs_{\pi}^{1} \to \PWs_{\pi}^{1}$ be a stable LTI system, $t \in \R$, and
  $f \in \PWs_{\pi}^{1}$. 
  There exists a monotonically increasing subsequence $\{N_k = N_k(t,f,T) \}_{k \in \N}$
  of the natural numbers such that
  \begin{equation*}
    \lim_{k \rightarrow \infty}
    (T_{N_k} f)(t)
    =
    (Tf)(t) .
  \end{equation*}
\end{theorem}
Theorem~\ref{th:convergent_subsequence} immediately implies the following corollary about
strong divergence.
\begin{corollary}
  For fixed $t \in \R$, all stable LTI systems
  $T \colon \PWs_{\pi}^{1} \to \PWs_{\pi}^{1}$, and all $f \in \PWs_{\pi}^{1}$ strong
  divergence of $(T_N f)(t)$ is not possible.
\end{corollary}

\begin{IEEEproof}[Proof of Theorem~\ref{th:convergent_subsequence}]
  Let $T \colon \PWs_{\pi}^{1} \to \PWs_{\pi}^{1}$ be a stable LTI systems, $t \in \R$,
  and $f \in \PWs_{\pi}^{1}$, all arbitrary but fixed. 
  To simplify the presentation of the proof, we assume that $f$ and $h_T$ are real valued. 
  If this is not the case, the following calculations need to be done separately for the
  real and imaginary part.

  From Theorem~\ref{th:uniform_convergence_arith_mean} we already know that
  \begin{equation}\label{eq:th:convergent_subsequence:arith_mean}
    \frac{1}{M} \sum_{N=0}^{M-1} (T_N f)(t)
  \end{equation}
  converges as $M$ tends to infinity, and that the limit is $(Tf)(t)$. 
  We distinguish two cases: first, the sequence $\{(T_N f)(t)\}_{N \in \N}$ converges
  itself, and second, $\{(T_N f)(t)\}_{N \in \N}$ diverges.

  We begin with the first case. 
  If $\{(T_N f)(t)\}_{N \in \N}$ converges then it converges to the same limit as
  \eqref{eq:th:convergent_subsequence:arith_mean}, which is $(Tf)(t)$. 
  In this case the proof is already finished.

  Now we treat the second case. 
  We assume that $\{(T_N f)(t)\}_{N \in \N}$ is divergent. 
  Then there exist two extended real numbers $a$ and $A$ ($a=-\infty$ and $A=\infty$ are
  possible) such that
  \begin{equation*}
    \liminf_{N \rightarrow \infty} (T_N f)(t) = a
  \end{equation*}
  and
  \begin{equation*}
    \limsup_{N \rightarrow \infty} (T_N f)(t) = A .
  \end{equation*}
  Note that we have $a<A$ due to the assumed divergence of $\{(T_N f)(t)\}_{n \in \N}$ and
  the convergence of the Ces{\`a}ro means \eqref{eq:th:convergent_subsequence:arith_mean}.
  
  Next, we show that
  \begin{equation}\label{eq:a_leq_Tf_leq_A}
    a \leq (Tf)(t) \leq A .
  \end{equation}
  If $a=-\infty$ or $A=\infty$ then the corresponding inequality in
  \eqref{eq:a_leq_Tf_leq_A} is trivially fulfilled. 
  Hence, we only have to show \eqref{eq:a_leq_Tf_leq_A} for $a>-\infty$ and $A<\infty$.
  Let $\epsilon > 0$ be arbitrary. 
  There exists a natural number $N_0=N_0(\epsilon)$ such that
  \begin{equation*}
    (T_N f)(t) > a - \epsilon
  \end{equation*}
  and
  \begin{equation*}
    (T_N f)(t) < A + \epsilon
  \end{equation*}
  for all $N \geq N_0$.
  Thus, we have for $M>N_0$ that
  \begin{align*}
    \frac{1}{M} \sum_{N=0}^{M-1} (T_N f)(t)
    &=
    \frac{1}{M} \sum_{N=0}^{N_0-1} (T_N f)(t)
    +
    \frac{1}{M} \sum_{N=N_0}^{M-1} (T_N f)(t) \\
    &>
    \frac{1}{M} \sum_{N=0}^{N_0-1} (T_N f)(t)
    +
    \frac{M-N_0}{M} (a-\epsilon)
  \end{align*}
  and
  \begin{align*}
    \frac{1}{M} \sum_{N=0}^{M-1} (T_N f)(t)
    &<
    \frac{1}{M} \sum_{N=0}^{N_0-1} (T_N f)(t)
    +
    \frac{M-N_0}{M} (A+\epsilon) .
  \end{align*}
  Taking the limit $M \rightarrow \infty$ yields
  \begin{equation*}
    (a-\epsilon) \leq (Tf)(t) \leq (A+\epsilon) .
  \end{equation*}
  Since this relation is true for all $\epsilon>0$, we have proved \eqref{eq:a_leq_Tf_leq_A}.

  Further, we have
  \begin{equation*}
    (T_N f)(t) - (T_{N-1} f)(t)
    =
    f(N) h_T(t-N) + f(-N) h_T(t+N)
  \end{equation*}
  which implies
  \begin{equation*}
    \lim_{N \rightarrow \infty} (T_N f)(t) - (T_{N-1} f)(t) = 0
  \end{equation*}
  by the Riemann--Lebesgue lemma \cite[p. 105]{grafakos08_book}.

  Next, we show that for every $L>0$ and $\mu>0$ there exists a natural number $\hat{N}$
  with $\hat{N} > L$, such that
  \begin{equation*}
    (T_{\hat{N}} f)(t) \in [(Tf)(t)-2\mu, (Tf)(t)+2\mu] . 
  \end{equation*}
  This shows that we can find a monotonically increasing sequence
  $\{\hat{N}_k\}_{k \in \N}$ such that
  \begin{equation*}
    \lim_{k \rightarrow \infty} (T_{\hat{N}_k} f)(t)
    =
    (Tf)(t) ,
  \end{equation*}
  and thus completes the proof. 
  Let $\mu > 0$ and $L>0$ be arbitrary but fixed. 
  We have to distinguish four cases: 1) $a>-\infty$ and $A<\infty$, 2) $a>-\infty$ and
  $A=\infty$, 3) $a=-\infty$ and $A<\infty$, and 4) $a=-\infty$ and $A=\infty$.

  We start with case 1).
  There exists a natural number $N_1 = N_1(\mu)
  > L$ such that
  \begin{equation*}
    \lvert (T_{N_1} f)(t) - a \rvert \leq \frac{\mu}{2} 
  \end{equation*}
  and
  \begin{equation}\label{eq:TN_minus_TNminus1_leq_mu_half}
    \lvert (T_{N} f)(t) - (T_{N-1} f)(t) \rvert \leq \frac{\mu}{2}
  \end{equation}
  for all $N > N_1$. 
  Further, there exists a natural number $N_2 > N_1$ such that
  \begin{equation}\label{eq:TN2-A_leq_mu_half}
    \lvert (T_{N_2} f)(t) - A \rvert \leq \frac{\mu}{2} .
  \end{equation}
  Let $\tilde{R}$ be the smallest natural number such that
  \begin{equation*}
    a + \tilde{R} \mu
    \geq
    A .
  \end{equation*}
  If $\tilde{R} = 1$, we have $A-a \leq \mu$, which implies that
  $(T_{N_1}f)(t) \in [(Tf)(t)-2\mu, (Tf)(t)+2\mu]$, and the proof is complete. 
  Hence, we assume $\tilde{R} \geq 2$. 
  For $n \in [0, N_2-N_1]$ we analyze $(T_{N_1 + n}f)(t)$. 
  Since we have \eqref{eq:TN_minus_TNminus1_leq_mu_half} and \eqref{eq:TN2-A_leq_mu_half},
  and $\tilde{R} \geq 2$, it follows that there exists at least one index
  $n_1=n_1(N_1, N_2, \mu) \in [1, N_2-N_1]$ such that
  \begin{equation}\label{eq:n1_in_interval1}
    (T_{N_1+n_1}f)(t) \in \Bigl(a + \frac{\mu}{2}, a+\frac{3}{2}\mu \Bigr] .
  \end{equation}
  We chose the smallest of these $n_1$, if there exist more than one. 
  If $\tilde{R}=2$, we stop. 
  If $\tilde{R} \geq 3$ we continue. 
  Due to \eqref{eq:TN_minus_TNminus1_leq_mu_half}, \eqref{eq:TN2-A_leq_mu_half}, and
  \eqref{eq:n1_in_interval1} there exists at least one index
  $n_2=n_2(N_1, N_2, \mu) \in [n_1, N_2-N_1]$ such that
  \begin{equation*}
    (T_{N_1+n_2}f)(t) \in \Bigl(a+\frac{3}{2}\mu, a+\frac{5}{2}\mu \Bigr] .
  \end{equation*}
  We chose the smallest $n_2$, if there exist more than one. 
  We continue this procedure until we have constructed the numbers $n_1(N_1, N_2, \mu)$,
  $n_2(N_1, N_2, \mu)$, \dots, $n_{\tilde{R}-1}(N_1, N_2, \mu)$. 
  Further, since $a \leq (Tf)(t) \leq A$, there exists exactly one natural number $r^*$
  with $0 \leq r^* \leq \tilde{R}$ such that
  \begin{equation*}
  (Tf)(t)
  \in
  \Bigl(a+\frac{2r^*-1}{2}\mu, a+\frac{2r^*+1}{2}\mu \Bigr] .
  \end{equation*}
  It follows that
  \begin{equation*}
    \lvert (T_{N_1+n_{r^*}} f)(t) - (Tf)(t) \rvert
    \leq
    \mu ,
  \end{equation*}
  which completes the proof for case 1).

  Next, we treat case 2). 
  Here have $-\infty < (Tf)(t) \leq A$. 
  We choose an arbitrary finite number $M$ such that $M<(Tf)(t)$. 
  It follows that $M < (Tf)(t) \leq A$. 
  Let $N_1$ be the smallest natural number such that $N_1=N_1(\mu)>L$,
  \begin{equation*}
    (T_{N_1} f)(t) \leq M ,
  \end{equation*}
  and
  \begin{equation*}
    \lvert (T_{N} f)(t) - (T_{N-1} f)(t) \rvert \leq \frac{\mu}{2}
  \end{equation*}
  for all $N \geq N_1$. 
  Now, we execute the same calculation as in case 1), where we replace $a$ by
  $a'=(T_{N_1}f)(t)$. 
  This completes case 2). 
  Case 3) is done analogously to case 2).
  
  In case 4) we have $a=-\infty$ and $A=\infty$. 
  We choose two arbitrary finite numbers $M_1$ and $M_2$ such that $M_1<(Tf)(t)<M_2$. 
  Let $N_1$ be the smallest natural number such that $N_1=N_1(\mu)>L$,
  \begin{equation*}
    (T_{N_1} f)(t) \leq M_1 ,
  \end{equation*}
  and
  \begin{equation*}
    \lvert (T_{N} f)(t) - (T_{N-1} f)(t) \rvert \leq \frac{\mu}{2}
  \end{equation*}
  for all $N \geq N_1$; and let 
  $N_2>N_1$ be the smallest natural number such that
  \begin{equation*}
    (T_{N_2} f)(t) \leq M_2 ,
  \end{equation*}
  and
  \begin{equation*}
    \lvert (T_{N} f)(t) - (T_{N-1} f)(t) \rvert \leq \frac{\mu}{2}
  \end{equation*}
  Now, we execute the same calculation as in case 1), where we replace $a$ by
  $a'=(T_{N_1}f)(t)$ and $A$ by $A'=(T_{N_2}f)(t)$. 
  This completes case 4) and thus the whole proof.
\end{IEEEproof}

\begin{remark}
  The proof of Theorem~\ref{th:convergent_subsequence} shows that in the case where
  $(T_N f)(t)$ is divergent, there exists for every real number $\xi \in [a,A]$ a
  monotonically increasing subsequence $\{\hat{N}_k(\xi)\}_{k \in \N}$ of the natural
  numbers such that
  \begin{equation*}
    \lim_{k \rightarrow \infty} (T_{\hat{N}_k(\xi)} f)(t)
    =
    \xi .
  \end{equation*}
\end{remark}

\section{Behavior of the Threshold Operator}

The threshold operator, which is of importance in many applications, maps all values below
some threshold to zero. 
If applied to the samples of the Shannon sampling series, the series becomes
\begin{equation}\label{eq:shannon_threshold}
  (A_{\delta} f)(t)
  \defequal
  \sum_{\lvert f(k) \rvert \geq \delta} f(k) \frac{\sin(\pi(t-k))}{\pi(t-k)} .
\end{equation}
In \eqref{eq:shannon_threshold}, only samples that are larger than or equal to the
threshold $\delta$ are considered. 
For $f \in \PWs_{\pi}^{1}$ and fixed $\delta>0$, the sum in \eqref{eq:shannon_threshold}
has only finitely many summands, because
$\lim_{\lvert t \rvert \rightarrow \infty} f(t) = 0$, according to the lemma of
Riemann--Lebesgue.

Like for the Shannon sampling series $S_N f$, where the truncation is done by considering
only the samples $f(k)$ where $\lvert k \rvert \leq N$, the convergence behavior of
$A_{\delta} f$ is of interest, as more and more samples are used in the sum, i.e., as
$\delta$ tends to zero. 
It has been shown that $A_{\delta} f$ is not globally uniformly convergent for
$\PWs_{\pi}^{1}$ in general.

In this paper, we analyze the behavior of the Hilbert transform of
\eqref{eq:shannon_threshold}, which is given by
\begin{equation*}
  (\Ht{A}_\delta f)(t)
  \defequal
  (\Hto A_\delta f)(t)
  =
  \sum_{\lvert f(k) \rvert \geq \delta} f(k) \frac{1-\cos(\pi(t-k))}{\pi(t-k)} ,
\end{equation*}
for functions $f$ in $\PWs_{\pi}^{1}$.

\begin{theorem}
  There exist a function $f_1 \in \PWs_{\pi}^{1}$ such that
  \begin{equation*}
    \lim_{\delta \rightarrow 0}
    \lVert \Ht{A}_\delta f_1 \rVert_{\infty}
    =
    \infty .
  \end{equation*}
\end{theorem}

\begin{IEEEproof}
  We use the function $f_1$ from the proof of
  Theorem~\ref{th:divergence_speed_conj_shannon_no_oversampl}. 
  We have $f_1(k) \geq 0$ and $f(k) = f(-k)$ for all $k \in \Z$, as well as
  $f_1(k) \geq f_1(k+1)$ for $k \geq 0$ and $f_1(k-1) \leq f_1(k)$ for $k \leq 0$. 
  Thus, for every $\delta$ with $0<\delta < f_1(0)$ there exists a natural number
  $N=N(\delta)$ such that
  \begin{equation*}
    (\Ht{A}_\delta f_1)(t)
    =
    \sum_{k = -N(\delta)}^{N(\delta)} f_1(k) \frac{1-\cos(\pi(t-k))}{\pi(t-k)} .
  \end{equation*}
  Due to the properties of $f_1$ we have $\lim_{\delta \rightarrow 0} N(\delta) = \infty$.
  This is a fixed subsequence. 
  According to the strong divergence, we have divergence for every subsequence.
\end{IEEEproof}

\section{Discussion}\label{sec:discussion}

\subsection{Divergence for Subsequences and Strong Divergence}\label{sec:discussion1}

Next, we treat question 4 from Section~\ref{sec:system-approximation}.

It is possible to state an approximation process for the Hilbert transform for which the
answer to Question Q2 is negative but the answer to Question Q1 is positive. 
For this approximation process, the question raised by Paul Erd{\H{o}}s in \cite{erdos41}
is to be answered negatively.

Let $f$ be a continuous $2\pi$-periodic function and $\Ht{f} \defequal \Hto f$ the Hilbert
transform of $f$. 
We only consider such $f$ for which $\Ht{f}$ is also continuous \cite{boche08e}. 
Equipped with the norm
$\lVert f \rVert_{B} = \lVert f \rVert_{\infty} + \lVert \Ht{f} \rVert_{\infty}$, this
space is a Banach space, which we denote by $B$.
We would like to approximate functions $f \in B$ by their finite Fourier series
\begin{equation*}
  (U_N f)(t)
  \defequal
  \frac{a_0}{2} + \sum_{k=1}^{N} (a_k \cos(kt) + b_k \sin(kt)) .
\end{equation*}
Then the Hilbert transform of $U_N f$ is given by
\begin{equation*}
  (\Ht{U}_N f)(t)
  \defequal
  \sum_{k=1}^{N} (a_k \sin(kt) - b_k \cos(kt)) .
\end{equation*}
We have
\begin{equation}\label{eq:fourier_series}
  (U_N f)(t)
  =
  \frac{1}{\pi} \int_{-\pi}^{\pi} f(\tau) D_N(t-\tau) \di{\tau}
\end{equation}
and
\begin{equation}\label{eq:conjugated_fourier_series}
  (\Ht{U}_N f)(t)
  =
  \frac{1}{\pi} \int_{-\pi}^{\pi} f(\tau) \Ht{D}_N(t-\tau) \di{\tau} ,
\end{equation}
where $D_N$ denotes the Dirichlet kernel, and $\Ht{D}_N$ is given by
\begin{equation*}
  \Ht{D}_N(t)
  =
  \frac{\cos \left( \frac{t}{2} \right) - \cos \left( \left( N + \frac{1}{2} \right) t
    \right)}{\sin \left( \frac{t}{2} \right)} .
\end{equation*}
For details, see for example \cite[p~59]{zygmund93_vol1_book}.

For every subsequence $\{N_k\}_{k \in \N}$ there exists a function $f_1 \in B$ such that
\begin{equation*}
  \limsup_{k \rightarrow \infty} \left| (\Ht{U}_{N_k} f_1)(t) \right|
  =
  \infty .
\end{equation*}
This follows directly from $\lim_{N \rightarrow \infty} \lVert U_{N,t} \rVert = \infty$,
where $U_{N,t} f \defequal (U_N f)(t)$, and the uniform boundedness theorem as discussed
in Section~\ref{sec:adapt-funct-reconstr}. 
However, we do not have strong divergence in this case. 
Because of \eqref{eq:fourier_series} and \eqref{eq:conjugated_fourier_series} we have
$\Ht{U}_N f= U_N \Ht{f}$. 
Since $\Ht{f}$ is also continuous, there exists, for every $f$ in $B$ and every
$t \in [-\pi,\pi)$, a subsequence $\{N_k\}_{k \in \N} = \{N_k(f,t)\}_{k \in \N}$ such that
\begin{equation*}
  \lim_{k \rightarrow \infty} (\Ht{U}_{N_k} f)(t)
  =
  \Ht{f}(t) ,
\end{equation*}
according to Fej{\'e}r's theorem \cite{fejer06}.

This is an example where the set of functions for which we have weak divergence is a
residual set, but where the set of functions for which we have strong divergence is empty,
i.e., an example where the uncountable intersection of residual sets is empty. 
This possibility was discussed in Section~\ref{sec:adapt-funct-reconstr}.

\subsection{Strong Divergence for Residual Sets}\label{sec:strong-diverg-resid}

In the following we want to gain a better understanding of question 5 in
Section~\ref{sec:system-approximation} by giving two examples in which we have strong
divergence for all functions from a residual set. 
It is important to note that the general behavior for strong divergence is unknown. 
In particular, it is unclear if for strong divergence we can have a similar situation as
in the Banach--Steinhaus theorem, where weak divergence for one function implies weak
divergence for all functions from a residual set. 
In order to obtain the results in this section we use very particular properties of
harmonic functions.

In the first example we consider the Hardy space $H^2$ and the quantity of interest is
$\max_{\omega \in [-\pi,\pi)} \lvert f(r \e^{\iu \omega}) \rvert$ as $r$ tends to $1$. 
The space $H^2$ consists of all holomorphic functions $f$ on the open unit disk $D$
satisfying
\begin{align*}
  \lVert f \rVert_{H^2}
  \defequal
  \sup_{0 < r < 1} \left( \frac{1}{2\pi} \int_{-\pi}^{\pi} \lvert f(r \e^{\iu \omega})
  \rvert^2 \di{\omega} \right)^{\frac{1}{2}}
  <
  \infty .
\end{align*}
For $0<r<1$ we define
\begin{equation*}
  M_r(f)
  \defequal
  \max_{\omega \in [-\pi,\pi)} \lvert f(r \e^{\iu \omega}) \rvert .
\end{equation*}
We have
\begin{equation*}
  \lVert M_r \rVert
  =
  \sup_{\lVert f \rVert_{H^2} \leq 1} M_r(f)
  \geq
  C_{3}(\epsilon) \sum_{n=1}^{\infty} \frac{1}{n^{1/2+\epsilon}} r^n ,
\end{equation*}
where
\begin{equation*}
  C_{3}(\epsilon)
  =
  \left( \frac{1}{\sum_{n=1}^{\infty} \frac{1}{n^{1+2\epsilon}}}\right)^{\frac{1}{2}} .
\end{equation*}
It follows that
\begin{equation*}
  \liminf_{r \rightarrow 1} \lVert M_r \rVert
  \geq
  C_{3}(\epsilon) \sum_{n=1}^{M} \frac{1}{n^{1/2+\epsilon}}
\end{equation*}
for all $M \in \N$, and consequently
\begin{equation*}
  \lim_{r \rightarrow 1} \lVert M_r \rVert
  =
  \infty .
\end{equation*}
Thus, the set of functions $f \in H^2$ for which we have
\begin{equation}\label{eq:limsup_Mr_inf}
  \limsup_{r \rightarrow 1} M_r(f)
  =
  \infty 
\end{equation}
is a residual set $\mathcal{D}$. 
Let $f_1 \in H^2$ be an arbitrary function satisfying \eqref{eq:limsup_Mr_inf}. 
According to the maximum modulus principle, we have for $0 < r_1 < r_2 < 1$, that
\begin{equation*}
  M_{r_1}(f_1)
  \leq
  M_{r_2}(f_1) .
\end{equation*}
Hence, we have
$\lim_{r \rightarrow 1} M_r(f_1) = \limsup_{r \rightarrow 1} M_r(f_1) = \infty$. 
This shows that we have strong divergence for all functions in the residual set
$\mathcal{D}$.

In the second example we consider the space $C(\partial D)$ of continuous functions on
$\partial D$, and the quantity of interest is
$\max_{\omega \in [-\pi,\pi)} \lvert (H_{\epsilon} f)(\e^{\iu \omega}) \rvert$ as
$\epsilon$ tends to $0$. 
$H_\epsilon$ is defined by
\begin{equation*}
  (H_{\epsilon} f)(\e^{\iu \omega})
  \defequal
  \frac{1}{2\pi} \int_{\epsilon < \lvert \omega_1-\omega \rvert \leq \pi}
  \frac{f(\e^{\iu \omega_1})}{\tan(\frac{\omega - \omega_1}{2})} \di{\omega_1} .
\end{equation*}
Let
\begin{equation*}
  u(r, \omega)
  \defequal
  \frac{1}{2 \pi} \int_{-\pi}^{\pi} f(\e^{\iu \omega_1}) \frac{1-r^2}{1-2r
    \cos(\omega-\omega_1)+r^2} \di{\omega_1}
\end{equation*}
denote the Poisson integral and
\begin{equation*}
  v(r, \omega)
  \defequal
  \frac{1}{2 \pi} \int_{-\pi}^{\pi} f(\e^{\iu \omega_1}) \frac{r \sin(\omega - \omega_1)}{1-2r
    \cos(\omega-\omega_1)+r^2} \di{\omega_1}
\end{equation*}
the conjugate Poisson integral \cite{zygmund93_vol1_book}. 
There exists a constant $C_{4}$ such that
\begin{equation*}
  \lvert (H_{\epsilon} f)(\e^{\iu \omega}) - v(1-\epsilon, \omega) \rvert
  \leq
  C_{4} \lVert f \rVert_{C(\partial D)}
\end{equation*}
for all $f \in C(\partial D)$, where
$\lVert f \rVert_{C(\partial D)} = \max_{\omega \in [-\pi,\pi)} \lvert f(\e^{\iu \omega})
\rvert$. 
Thus, it follows that
\begin{equation*}
  \left| \max_{\omega \in [-\pi,\pi)} \lvert (H_{\epsilon} f)(\e^{\iu \omega}) \rvert -
    \max_{\omega \in [-\pi,\pi)} \lvert v(1-\epsilon, \omega) \rvert \right|
  \leq
  C_{4} \lVert f \rVert_{C(\partial D)}
\end{equation*}
for a universal constant $C_{4}$. 
Let $\omega \in [-\pi, \pi)$ be fixed. 
Then the set of all functions $f \in C(\partial D)$ with
\begin{equation*}
\limsup_{r \rightarrow 1} \lvert v(r,\omega) \rvert
=
\infty  
\end{equation*}
is a residual set \cite{zygmund93_vol1_book}. 
According to the maximum principle for harmonic functions, it follows that
\begin{equation}\label{eq:lim_max_v}
  \lim_{r \rightarrow 1} \max_{\omega \in [-\pi,\pi)} \lvert v(r,\omega) \rvert
  =
  \infty .
\end{equation}
Since the set of function which satisfies \eqref{eq:lim_max_v} is a residual set, it
follows that the set of functions $f \in C(\partial D)$ with
\begin{equation*}
  \lim_{\epsilon \rightarrow 0} \max_{\omega \in [-\pi,\pi)} \lvert (H_{\epsilon}
  f)(\e^{\iu \omega}) \rvert
  =
  \infty
\end{equation*}
is a residual set.

\begin{remark}~
  \begin{enumerate}
  \item For functions $f$ in the Hardy space $H^2$, which was discussed in the first example,
    the Poisson integral converges for $r \to 1$ in the $L^2$-norm to the function
    $f(\e^{\iu \omega})$, $\omega \in [-\pi,\pi)$, i.e., we have
    \begin{equation*}
      \lim_{r \rightarrow 1} \frac{1}{2\pi} \int_{-\pi}^{\pi} \lvert u(r,\omega) -
      f(\e^{\iu \omega}) \rvert^2 \di{\omega}
      =
      0 .
    \end{equation*}
    The point evaluation operator of $f$, which maps $H^2$ to $\C$ and which is defined by
    $f \mapsto f(\e^{\iu \omega})$, is unbounded (and only well-defined except for sets of
    Lebesgue measure equal to zero).
  \item In the second example above, we discussed the Hilbert transform. 
    For $f \in C(\partial D)$, the Hilbert transform $\Hto f$ is defined as a
    $L^2$-function. 
    In general the Hilbert transform is not bounded, i.e., we have
    \begin{equation*}
      \sup_{\substack{\lVert f \rVert_{C(\partial D)} \leq 1 \\ f \in C^{\infty}(\partial D)}} \lVert \Hto f \rVert_{C(\partial D)}
      =
      \infty ,
    \end{equation*}
    where $C^{\infty}(\partial D)$ denotes the set of infinitely often differentiable
    functions on $\partial D$.
  \end{enumerate}
  In both examples, the unboundedness of the operators enables us to show strong
  divergence on a residual set.
  In further studies \cite{boche15c_accepted}, it became clear that the unboundedness of
  the operators is necessary to have strong divergence for a residual set.
\end{remark}

\subsection{Final Remarks and Future Work}

We have shown that for the Shannon sampling series, the conjugated Shannon sampling
series, and for more general system approximation processes based on equidistant sampling
we can have strong divergence. 
Further, oversampling does not improve this behavior in general. 
This answers question 3 from Section~\ref{sec:system-approximation}. 
We have also shown that for pointwise system approximation, strong divergence cannot
occur.

For the approximation of the Hilbert transform of continuous $2\pi$-periodic functions
with continuous Hilbert transform the Question Q2 in
Section~\ref{sec:adapt-funct-reconstr} has to be answered negatively and Question Q1
positively. 
Moreover, for the Hilbert transform we can have strong divergence for a residual set. 
In all constructions and examples we use specific properties of the underlying function
spaces and systems.

It would be interesting to develop a general theory for strong divergence, in particular
because such a theory can constitute the basis of an adaptive signal processing approach,
as it was discussed in Sections \ref{sec:adapt-funct-reconstr} and
\ref{sec:system-approximation}.

Recently, a first step toward this general theory was made in \cite{boche15c_accepted}.
As we already pointed out, the unboundedness of the operators in
Section~\ref{sec:strong-diverg-resid} is responsible for having strong divergence of the
approximation processes on a residual set.
In \cite{boche15c_accepted} it was shown that for the approximation of bounded operators,
strong divergence can occur at most on a meager set, and not on a residual set.
The operators associated with the Shannon sampling series, the conjugated Shannon sampling
series, and the system approximation process, i.e., the identity, the Hilbert transform
and the LTI system under consideration, are bounded operators for the Paley-Wiener space
$\PWs_{\pi}^{1}$.
Therefore, the divergence behavior from Sections
\ref{sec:behav-conj-shann} and \ref{sec:overs-with-kernels} can only occur for functions
from a meager set.

\section{Acknowledgments}

The authors thank Ingrid Daubechies for valuable discussions of questions Q1 and Q2
especially on the construction of good reconstruction kernels at Strobl 2011 and at the
``Applied Harmonic and Sparse Approximation'' workshop at Oberwolfach in 2012. 
The first author thanks Rudolf Mathar for his insistence in several conversations on the
importance of understanding the strong divergence behavior addressed here. 
The authors also thank the referees of the German Research Foundation (DFG) grant BO
1734/13-2 for highlighting these questions as well in their review. 
Further, thanks are to Ezra Tampubolon for carefully reading the manuscript and providing
helpful comments.


\end{document}